\documentclass[preprint]{aastex}

\slugcomment{To Appear in the December 2000 Astronomical Journal}

\shorttitle{Circumstellar Disks and Protostars in Orion}
\shortauthors{Lada et al.}

\begin{document}


\title{Infrared L Band Observations of the Trapezium Cluster:\\
    A Census of Circumstellar Disks and Candidate Protostars}


\author{Charles J. Lada}
\affil{Harvard-Smithsonian Center for Astrophysics, 60 Garden Street,
    Cambridge, MA 02138}
\author{August A. Muench\altaffilmark{1}}
\affil{Department of Astronomy, University of Florida, Gainesville, FL 32611
and\\ 
Harvard-Smithsonian Center for Astrophysics, Cambridge, MA 02138}
\author{Karl E. Haisch Jr.} 
\affil{Department of Astronomy, University of Florida, Gainesville, FL 32611}
\author{Elizabeth A. Lada}
\affil{Department of Astronomy, University of Florida, Gainesville, FL
32611}
\author{Jo\~{a}o F. Alves}
\affil{European Southern Observatory, Karl-Schwartzschild-Strasse 2,
85748 Garching Germany}


\author{Eric V. Tollestrup\altaffilmark{2} \& S. P. Willner}
\affil{Harvard-Smithsonian Center for Astrophysics, 60 Garden Street,
Cambridge, MA 02138}


\altaffiltext{1}{Smithsonian Predoctoral Fellow}
\altaffiltext{2}{present address:
Department of Astronomy, Boston University, 
Boston MA 02215}


\begin{abstract}
We report the results of a sensitive near-infrared JHKL imaging survey
of the Trapezium cluster in Orion. We use the JHKL colors to obtain 
a census of infrared excess stars in the cluster.
Of (391) stars brighter than 12th magnitude in the
K and L bands, 80\% $\pm$ 7\% are found to exhibit detectable infrared
excess on the J-H, K-L color-color diagram. Examination of
a subsample of 285 of these stars with published spectral types yields a
slightly higher infrared excess fraction of 85\%. 
We find that 97\% of the optical proplyds in the cluster
exhibit excess in the JHKL color-color diagram indicating that 
the most likely origin of the observed infrared excesses is from  
circumstellar disks. We interpret these
results to indicate that the fraction of stars in the cluster with
circumstellar disks is between 80-85\%,
confirming earlier published
suggestions of a high disk fraction for this young cluster.
Moreover, we find that the probability of finding 
an infrared excess around a star is independent of
stellar mass over essentially 
the entire range of the stellar mass function down to the hydrogen
burning limit.
Consequently, the vast majority of stars
in the Trapezium cluster appear to have been born with circumstellar
disks and the potential to subsequently form planetary systems,
despite formation within the environment of a
rich and dense stellar cluster.
We identify 78 stars in our sample characterized by 
K-L colors suggestive of deeply embedded 
objects. The spatial distribution of these objects differs from
that of the rest of the cluster members and is similar to that of 
the dense molecular cloud ridge behind the cluster.
About half of these objects are detected in the short 
wavelength (J and H) bands and these are found to be characterized by
extreme infrared excess.  
This suggests that many of these sources could be protostellar in
nature. If even a modest fraction fraction (i.e., $\sim$ 50\%) 
these objects are protostars, then 
star formation could be continuing in the molecular ridge 
at a rate comparable to that which produced the foreground Trapezium cluster.

\end{abstract}


\keywords{clusters: galactic, circumstellar disks, Trapezium, star formation,
infrared: stars}


\section{Introduction}

The discovery of small amplitude Doppler periodicities
in the spectra of a number of nearby, sunlike, stars has established 
the existence of extra-solar planets beyond doubt  \citep{mb99}.  
However, we do not yet know how common planetary systems
are in the Galaxy. Obtaining a meaningful census via direct and indirect
detections of planetary systems is a formidable challenge which will 
likely require the development of new technological capabilities such as
precision photometric and astrometric observations from space. On the other
hand, an estimate of the frequency of planetary systems
is indirectly possible via observations of protoplanetary systems. 
Such observations are feasible with existing technology, even
from the ground. 

The low inclinations and eccentricities of the planets around the
sun have long suggested that the solar system formed in a 
disk of dust and gas which surrounded the sun in its earliest
stages of evolution. Moreover, theory suggests 
that circumstellar disk formation is the 
natural consequence of star formation in rotating collapsing
protostellar clouds \citep{sal87} and that protoplanetary
disks should be relatively common in the Galaxy. A protoplanetary disk
is considerably easier to detect than a planetary system with a  
similar mass of solid material \citep{bs96}. 
Indeed, circumstellar disks have been directly and indirectly 
observed around numerous young stellar objects 
via optical, infrared, centimeter, and millimeter-wave  observations 
\citep[e.g.,][]{svwb99,mab99,clada99,whkr00}
and even resolved in optical and millimeter-wavelength images
\citep{mo96,dutrey96}.  
Thus obtaining a statistically meaningful census of
circumstellar disks around young stars
may be a very efficient first step toward determining the 
the frequency of planetary systems in the Galaxy.

One particularly efficient method of producing meaningful
measurements of disk frequencies around young stars is to  
obtain multi-wavelength, infrared imaging surveys of young 
stellar clusters.  Circumstellar disks are bright infrared
emitters and produce emission in excess of that emitted by
the photospheres of young stars. Such infrared excess is
readily measured with near-infrared imaging observations.
Young clusters contain hundreds of young target stars
spanning nearly the full range of stellar mass within a 
relatively small angular extent on the sky. Infrared imaging surveys 
can thus obtain a complete sampling of cluster members in 
relatively modest observing times.

Infrared color-color diagrams, constructed from multi-wavelength imaging
surveys, have been shown to be a useful tool for identifying infrared
excesses and candidate circumstellar disks around stars
in young clusters \citep{la92}.  JHK color-color diagrams have been
the most commonly employed to investigate infrared excess and 
circumstellar disk frequency in young clusters. 
Such studies have produced evidence which suggests that the
frequency of infrared (JHK) excess sources 
decreases with cluster age on timescales on the order of only a
few million years, and this potentially sets important constraints
on both the lifetimes of circumstellar disks and the duration of
the planet building phase within them \citep{ll95,lal96,elada99}. 
However, JHK observations do not extend to long enough 
wavelength to enable either a complete or unambiguous census 
of circumstellar disks in young clusters for two reasons.
First, the magnitude of the near-infrared excess produced by a 
star/disk system depends on the parameters such as the disk
inclination, accretion rate and presence and size of 
inner disk holes \citep{als87,la92,mch97}, and not all
the star/disk systems in
a young stellar population are expected to exhibit large enough excesses
to be detected in a JHK color-color diagram \citep{la92,kh95,
hetal98}. Second, a JHK excess can be produced by other
phenomena such as extended emission from reflection nebulae and
H II regions, which are common in regions of recent star formation
(e.g., Haisch, Lada \& Lada 2000, hereafter HLL).

The magnitude of the infrared excess produced by a disk 
rapidly increases with wavelength, and 
L band (3.5 \micron )  
is likely the optimum wavelength for detecting infrared excesses
from protoplanetary disks with ground-based telescopes (HLL).
Existing imaging detectors are capable of detecting the photospheres
of relatively low mass stars in nearby regions, and
circumstellar disks produce sufficiently
strong L band excess that almost all stars that have disks can
be identified with JHKL (or HKL) observations, independent of 
inclination, disk accretion rates, etc. \citep{kh95}. 
Moreover, extraneous emission from reflection nebulae, 
HII regions, etc. is reduced at this wavelength. 

This paper is the second in a
series which investigate the frequency and
evolution of circumstellar disks using L band observations of 
nearby young clusters. The first paper (HLL) reported results of an L band survey
of NGC 2024, a heavily embedded and extremely young cluster 
in Orion. This paper reports the results of a JHK and L band imaging survey
of the famous Trapezium cluster also in Orion.

The Trapezium cluster is a very rich cluster within the Great Orion Nebula.
Its members provide a statistically meaningful 
sample of nearly the entire spread of the stellar mass function.
The cluster is somewhat older than NGC 2024 but not as embedded in molecular
material, and it is much better studied over a wide range of 
wavelength \citep[e.g.,][]{ht86,church87,felli93,mcc95,setal94,ad95,h97}.
There is ample observational evidence for a large population 
of star/disk systems in the Trapezium cluster. Indeed, 
the resolved HST images of silhouetted disks within the cluster 
\citep{mo96} are perhaps the most compelling demonstration
for the existence of circumstellar disks around young stars yet
obtained. However, the size of the disk population in this cluster
is uncertain.

Around the O star $\theta^1$C Ori, centimeter-wave
radio observations \citep{church87,felli93}
and optical imaging both from the ground \citep{
lv79} and with HST \citep{owh93} revealed what appear to be
numerous photo-evaporating disks in the inner region of the 
cluster. A detailed comparison of these data with deep, 
high resolution infrared (I \& K band) images suggested 
that the fraction of stars with disks in this part of 
the cluster is $>$ 25-50\% and could even be as high as
100\% \citep{setal94}. A detailed analysis of near-infrared
IJHK band observations of the larger ONC (Orion Nebula Cluster) by
\citet{hetal98}  suggested a similarly high disk fraction,
estimated to be 55-90\%. This estimate was 
supported indirectly by observations
of Ca II emission lines which suggested disk accretion occuring
in about 70\% of the stars in the larger cluster. Infrared
L band observations could both confirm and considerably improve
upon these estimates of disk frequencies. 
According to \citet{hetal98} only 43 of 1600 visible stars in the ONC cluster
have published  L measurements and more complete
L band observations 
``...would provide more robust estimates of disk frequencies
among both visible and optically obscured young stellar populations.'' 
Earlier, \citet{mrzs96} briefly noted an estimate of 60-80\%
for the Trapezium disk fraction based on a very rough analysis
of L band imaging observations,
suggesting that more detailed L band data and analysis could indeed
produce a relatively accurate disk census for the cluster.
The  L band imaging survey reported here confirms these 
expectations. We derive a significantly improved and
very high disk frequency of 80-85 \%,
for the Trapezium region.

The stars in the Trapezium cluster formed
at a relatively high rate (e.g., Lada, Alves \& Lada 1996).
The cluster is in very close proximity to, but not deeply embedded in, 
a dense molecular cloud. Its physical association 
with the molecular cloud raises the interesting issue
of the extent to which star formation is still continuing in the 
cluster region.  Recently, \citet{ps99} have presented evidence indicating 
that the star formation rate in this cluster has been accelerating
with time. However, they postulate a sharp decline in the present day star 
formation rate due to O star formation. Have most of the stars which will be 
members of this cluster already formed? Or is rapid
star formation continuing to add significant numbers to
the cluster population within the remaining molecular gas? 
Knowledge of the protostellar population in the cloud is
essential to address these questions. 
Protostars are expected to exhibit 
large L band excesses and very red colors indicative of deeply
embedded objects.
Our L band observations are both long enough in wavelength 
and sufficiently sensitive to
provide a preliminary reconnaissance of the protostellar population
in this cluster/cloud region.

Section 2 discusses the instrumentation
employed for this experiment and the data acquisition and
reduction procedures used. Section 3 describes the results
and analysis of our observations and Section 4 
discusses their implications.  Section 5 summarizes the
primary findings.

\section{Observations}

\subsection{Instrumentation}

Observations of the Trapezium cluster were
obtained at the Fred Lawrence Whipple
Observatory on Mt. Hopkins, Arizona, using the 1.2m (48in)
f/8 Ritchey-Chretien reflector telescope and
in conjunction with SAO's dual channel infrared camera, STELIRCAM
\citep{tw98}.
STELIRCAM consists of two 256$\times$256
InSb detector arrays. Each is fed from a dichroic mirror that separates
wavelengths longer and shorter than 1.9 $\mu$m into two independent imaging 
channels for simultaneous observations on
the sky.  Three separate magnifications can be selected by moving  
different cold lens assemblies into the beam. For our  JHKL
survey we selected a field of view of
2.56\arcmin$\times$2.56\arcmin \hspace*{0.05in}with a resolution of
0.6\arcsec \hspace*{0.05in}per pixel. The magnification is
the same for both arrays and both channels simultaneously view approximately
the same field on the sky.
For all of these observations, the J$_{\mbox{Barr}}(1.25\;\mu\mbox{m})$,
H$_{\mbox{Barr}}(1.65\;\mu\mbox{m})$,
K$_{\mbox{Barr}}(2.2\;\mu\mbox{m})$,
and L$_{\mbox{Barr}}(3.5\;\mu\mbox{m})$ filters were used.

\subsection{Data Acquisition and Reduction}

JHKL band images of the Trapezium cluster were obtained on 14, 15,
and 16 December 1997 and JHL images on 04 and 05 November 1998.
In both observing runs the cluster was imaged in a square mosaic pattern
approximately centered on the {\rm O}7 star HD 37022 ($\theta^1$C Orionis:
R.A.= 05 35 16.47; DEC.=-05 23 22.90; J2000).
The individual mosaic images consist of nine separate
STELIRCAM fields arranged in a three by three grid.
The 1997 fields were spatially overlapped by 34\arcsec\  ($\sim$ 25\% of the array)
in both right ascension and declination providing for accurate
positional placement of mosaiced fields 
as well as redundancy for the photometric
measurements of sources located in the overlapped regions.
The resulting mosaiced field  covered an area of $\sim$36 arcmin$^{2}$
(i.e., 6.55\arcmin\  by 6.55\arcmin\ ) but
because the two arrays were not exactly
coincident on the sky the actual area covered simultaneously
at all wavelengths was slightly smaller. Typically, the dual
channel camera was configured to obtain either simultaneous H and
K band or J and L band observations. 

Nine off-cluster control fields were also observed in 1997 December
to determine accurate sky and flat field calibration frames and measure the
colors of field stars.
The control fields for these observations were centered at 
R.A.= 05 26 00; DEC.=-06 00 J2000 and were determined to 
be free of molecular material by inspection of the Palomar
Sky Survey Prints and the 100 micron dust opacity maps
of \citet{wmd94}.  Our observing procedure was to intersperse
observations of control fields with observations of cluster
fields. Typically control field observations were obtained
after three consecutive positions in the cluster mosaic were
imaged. 

Each position in the mosaic was 
observed in a 9 point dither pattern with
small, 12\arcsec \ offsets between dithers.
The integration times and co-additions per dither
were 15 seconds with 4 co-additions for J, H and K bands, and 
0.1 seconds with 400 co-additions at L band, 
yielding total integration times of 9 minutes
for JHK and 6 minutes at L band.
Dark frames were observed at the beginning and end of each night,
using the same combinations of integration times and co-additions
as the observed images.

\cite{el82} infrared standard stars were observed throughout
the night at JHKL wavelengths. 
These included HD 203856, HD 225023, HD 18881, HD 40335, HD 44612, 
HD 84800, HD 105601. Magnitudes are reported in the instrumental
system of STELIRCAM.

Large temporal background variations at L band
(due to window dust, instrument/telescope flexture, 
and detector anomalies),
coupled with the relatively long integration
times employed and significant extended L band 
emission from the HII region made it difficult to determine accurate 
sky levels across the L band mosaic image. As a result
the 1997 data proved to be of relatively poor photometric
quality and could not be used to construct a mosaic
image of sufficient quality to provide a useful
image of extended nebulous emission.       
Therefore additional L band images were obtained in
the telescope nodding mode
on 4 and 5 November 1998 to improve the sky subtraction and
our ability to faithfully image the extended L band emission from
the Orion nebula.
In this paper we only use this latter
data set for the L band measurements.
Short integration J and H observations were simultaneously
obtained to extend the dynamic range of observations in those 
bands and improve the photometry of bright sources.

For each field in the 3 x 3 mosaic grid observed in November
1998 we obtained a single exposure
and then nodded the telescope to a sky position 450\arcsec
\hspace*{0.05in} west of the cluster where an identical exposure
was obtained. After repeating this sequence for all 9 positions in the grid,
the center of the grid was shifted by a small dither ($\sim$ 5-10\arcsec\
in a random direction) and the entire mosaic was observed again.
This was repeated a total of 18 times.
For 9 passes the camera was configured to acquire simultaneous J and L
band observations and for the remaining 9 passes H and L observations.
The integration times for a single position were 1.0 second with 
12 coadds at J and H bands and 0.1 seconds with 100 coadds at L band.
The total integration time for each field was 3 minutes in the L band.
Overlaps of 20\arcsec\  between positions in this mosaic provided a
total area of approximately 7.0\arcmin\  by 7.0\arcmin\  at
J,H and L bands, or
slightly larger than the 1997 observations.
These images were observed only near transit, with a range of 
airmass of only 1.25 to 1.28. However, since no L band standards
were obtained for the November observations, the L band 
photometry was calibrated internally using 34 stars whose
L band magnitudes were independently obtained from observations with the NASA
Infrared Telescope Facility (IRTF) on Manua Kea Hawaii in February 2000. The resulting 
uncertainty in the calibrated L photometry was 
0.05 magnitudes.  For 14 stars with 
previous L band photometry in the literature
the magnitudes agree with the 
calibrated November photometry to within 
the 0.05 magnitude calibration uncertainty.
Finally, the entire set of November 1998 photometry was
directly compared to the more noisy but independently calibrated
1997 December observations (for which simultaneous L band observations
of standards were obtained) and found to agree but within a
larger uncertainty due to the poor photometric
quality of the 1997 data set. 

All data were reduced using standard routines in
the Image Reduction and Analysis Facility
(IRAF) and the Interactive Data Language (IDL).
An average dark frame was constructed from the darks taken at
the beginning and end of each night's observations. This dark frame was
subtracted from all target 
observations to yield dark subtracted images. 
After appropriate linearization,  all dark
subtracted target frames were then processed by subtracting the
appropriate sky frame and dividing by the flat field.
For the 1997 observations, sky frames were constructed from 
the off-cluster control field images by median filtering the
nine dithered observations obtained at each off field position.
Flat field frames were constructed by normalizing the median
sky frames. For the nodded 1998 observations, sky frames were
constructed by median combining the nearest 9 sky positions in time to
the target observation. Each sky frame was checked to confirm that all
stars had been removed by this process. The sky frames were then
normalized to produce flat fields for each target frame.
The resulting frames were individually inspected to ensure that the best sky
subtraction and flat field had been used. Finally all target frames for
a given position in the cluster were registered and combined to produce
the final image of each field.

Sources were extracted from each of our images using the automated
source extractor DAOFIND in IRAF. The detection threshold was set
to be 5 $\sigma$ above the average sky noise in an image. Each image was then
inspected by eye to recover sources missed by DAOFIND. Photometry
was then obtained for all the identified sources using PSF fitting routines
in the DAOPHOT package \citep{Stetson87} in IRAF. The primary motivation
for using PSF fitting photometry was to mitigate the effect of small
scale variations in the background ``sky'' due to the extended emission
of the Orion nebula. This was made possible by use of special 
sky-fitting routines within the DAOPHOT package as briefly described below.

For the JHK images
twenty bright, unsaturated and isolated stars were selected in each
of the nine individual images for the construction of a master PSF for that
mosaic position.  For the L band 
observations a single PSF was constructed for the entire mosaic image
from 30 bright, unsaturated and isolated stars within the mosaic field.  
This was possible because the PSF did not vary across the mosaic 
due to the manner in which the observations were obtained. (As described
earlier, the entire 3 x 3 mosaic was observed in one pass 
with short (10 seconds) individual integration times. 
The final mosaic was constructed from median filtering 18 separate
passes.)  
Sky-fitting routines were employed to improve the subtraction of 
nebulous emission from the stellar profiles. These routines iteratively
obtain estimates for the local sky by removal of target stars and
re-measurement of the residual sky \citep{parker91}.
Artificial stars added to a very nebulous K band image 
showed a factor of two improvement in the photometric errors with 
the sky-fitting option.

All sources were placed in a single positional grid. Absolute
positions for our survey stars were established by 
comparison with the positions for 45 infrared
sources determined by K. Luhman (private communication).
The relative positional uncertainty in the derived coordinates
was found to be 0.2 arc seconds. The entire data base
of our survey observations, including complete photometry
and positions of all sources will be published in a separate
contribution \citep{gus2000}.

\section{Results and Analysis}

\subsection{Images}

Figure 1a shows our JHK mosaic image 
of the Trapezium cluster. This is a false color image with J band coded
as blue, H band as green and K band as red. The image is displayed in
the standard orientation with North at the top and West on the right.
The four famous Trapezium stars are in the center of the image. 
The distinctive reddish nebulosity northwest of the Trapezium stars
marks the location of the well known Kleinmann-Low/IRC2 protostellar region.
Figure 1b shows the L band mosaic image constructed
from our observations. 
Extended L band emission is evident throughout the
region with the well known bright bar prominent in the southeast.
The extended L band nebulosity is due to emission from
a prominent dust emission feature at 3.3 microns which originates from small
grains (PAHs) which are excited by UV radiation from 
the OB stars in the cluster \citep[e.g.,][]{sell81, dw81,geballe89}. 
Comparison of the two images shows a clear displacement
of the position of the bar in the L band image relative to the JHK
image. The L band emission is
located further downstream from the exciting Trapezium
stars than the J band emission which dominates the emission in the
JHK image. While the J band emission traces 
the ionized gas of the HII region, the L band emission appears to
originate within the photodissociation region at the interface
between the HII region and the molecular cloud from which
the cluster is emerging \citep{geballe89,roche89}. 

Another interesting extended feature apparent in the L band 
map is a long and thin comet-like structure which appears 
to emanate from one of the infrared sources approximately
56 arc seconds to the east and 14 arc seconds to the 
north of $\theta^1$C Ori. 
This source is located within a narrow dark filament
at the bottom edge of the well known ``dark bay'' which is
prominently seen in optical images (for example, see the HST WFPC image of
Orion Nebula). In Figure 2 we show a close up view  of this feature in
both the L and J bands. The narrow dark filament or lane
is clearly evident in the J band image, although there is
no trace of the cometary structure. The cometary feature
appears most clearly in the L band image and is extended
at a position angle of $313^o$ from the infrared source,
nearly orthogonal to the dark filament. The cometary tail
extends for roughly 21\arcsec \ or 9400 AU from the star
and is very red, hardly visible in the K or H band images.
Although morphologically similar to many of the proplyds
observed by HST in this nebula, the L band cometary feature
is considerably longer than any of the optical proplyds and moreover
does not point back to $\theta^1$ C Ori, the primary ionizing source
of the proplyd population. In addition the associated stellar
component is identified as a non-nebulous star in the
proplyd survey of O'Dell and Wong (1996), and it does not
appear as a radio source in the Felli et al. (1993 ) 
survey. On the other hand,
this star has a strong L band infrared excess indicative
of a circumstellar disk. This cometary feature could be a jet ejected
by the star/disk system at its apex, but its very red
color would be still difficult to explain. Our data do not allow
us to further constrain the nature of this interesting feature.
Narrow-band images (e.g., H$_2$) could reveal shock excited
emission within this object and test the interesting proposition that it
may be a jet.

Within the region surveyed, we detected above our 10 $\sigma$ detection
limits 548  L sources (m$_L < 12.0$), 
729  K sources (m$_K <$ 16.8), 695  H sources (m$_H <$ 17.7),
and 600  J sources (m$_J <$ 18.5). Of these 10 sigma detections, 595 were simultaneously
detected at  JHK, 490 simultaneously at  JHKL and 
46 only at  HK. Five sources were detected only at L band.
Overall a total of 734 distinct sources were detected.

\subsection{Color-Color Diagrams: The Infrared Excess Fraction}

Figures 3 and 4 show the JHK and JHKL color-color diagrams
for those (391) sources with both K and L 
$<$ 12 magnitudes. By selecting sources
in this way, our survey only includes those sources whose photospheres
are detectable at both K and L bands.
This insures a meaningful determination of the excess/disk fraction 
in this cluster. Almost 
all the stars with K $<$ 12 magnitudes have photometric uncertainties
less than 0.1 magnitudes in the J, H and K bands. 
The typical source in our sample is also characterized by
photometric uncertainties less than 0.2 magnitudes at L band. Furthermore,
we excluded from our sample any source with an uncertainty greater
than 0.25 magnitudes at L band. 

The solid curve is the locus of colors
corresponding to main sequence stars which range in spectral 
type from early O to M5 \citep{bb88}. The solid straight 
line extending from the main sequence curve is the locus of
colors of classical T Tauri stars \citep{mch97}.
The three dashed parallel lines
define the reddening bands for main sequence stars and
classical T Tauri stars (CTTS), respectively.
These lines are parallel to
the reddening vector given by the \citet{cfpe81} reddening
law which we have adopted for this study. 

All stars which fall to the right of the main
sequence reddening band possess infrared excess and are 
circumstellar disk candidates.
In the JHK diagram 50\%  $\pm$ 7\% of the stars are found to display colors 
indicative of infrared (K band) excess. In the JHKL diagram, 
80\% $\pm$ 7\% of the stars are found to have infrared (L band) excess.

The fraction of stars falling in the excess region of the color-color
diagram is sensitive to the adopted reddening law and to a lesser
extent to the photometric system used to derive the reddening law.
We have determined JHK and JHKL excess fractions for two other 
reddening laws (obtained in different photometric systems)
\citep{koor83,rl85} and find JHK excess fractions of 50\% and 49\% and
JHKL excess fractions of 78\% and 75\%, respectively. These values
are within the quoted (statistical) uncertainties of our determinations
using the \citet{cfpe81} extinction law in the STELIRCAM photometric 
system.

\subsection{The K-L Distribution: Candidate Protostars}

Many of the reddest and most heavily extincted stars will not 
be detectable at J or H band and will not appear in the color-color
diagrams. Therefore it is useful to examine the distribution of K-L
colors of the stars in our survey. Figure 5 displays the frequency
distribution of K-L color for the 603 stars in our sample detected in 
both bands. Fifty-seven  of these objects were not detected at either J
or H. The distribution of the 603 stars detected in both bands
has a broad but prominent peak at K-L $\approx$ 0.9
magnitudes. This distribution is very similar to that in NGC 2024
derived by HLL. There is no peak near zero color
similar to that found by \citet{kh95} for YSOs in Taurus and
characteristic of non-excess or diskless (Class III or WTTS) stars.
The shift of the peak to relatively large
K-L color is primarily due to infrared 
excess emission and to a lesser extent to reddening.
It would require about 25 magnitudes of visual
extinction to produce a K-L color of 1.0 magnitude in a naked, diskless
star but examination of the color-color diagrams 
shows that the vast majority of stars in the cluster have
extinctions well under 10 visual magnitudes. 
Thus, for the Trapezium cluster, the presence of infrared
excess emission is the primary cause of this shift to larger values
of the K-L color.

In the Taurus population of YSOs, objects with K-L $>$ 1.5
magnitudes are almost always protostellar in nature \citep{kh95}.
In order to compile a catalog of potential protostellar objects 
in the Trapezium region we consider those sources in our survey
with similarly large  K-L colors.
In the Trapezium we find 94 stars with K-L $>$ 1.5.
Of these, 59 (63\%) are also detected in the 
J and H bands. Twenty-one of these latter objects possess
anomalous colors in the JHKL color-color diagram (see
discussion in 3.4 below), likely due to source variability, and/or photometry
contaminated by unresolved stellar pairs. These sources are not
likely to be protostellar objects. However, 
28 of the 38 non-anomalous very red sources detected in all four bands
are located to the right of the CTTS 
reddening band above the termination of the (unreddened) CTTS locus.
Sources in this part of the color-color diagram possess extreme 
excess emission which is often characteristic of protostellar (Class
I; see \citet{clada99}) objects.
Such sources are frequently embedded in bright infrared
reflection nebulae which can shift otherwise heavily reddened sources into
this region of color-color diagram by depressing the J-H color
relative to the K-L color. Sources in this region therefore
can be appropriately considered as
candidate protostars.  The remaining 10 very red sources detected
in all bands are located for the most part 
high within the CTTS reddening band. The colors of these
sources are also consistent with those of protostellar objects, and they
can also be considered as reasonable candidates for protostellar
status. There are 35 sources with K-L $>$ 1.5 that were not detected
in either J or H band (and cannot be placed on the color-
color diagram). These are also 
candidate protostars.  Although some of these sources
could be non-protostellar objects that are more embedded than
stars of similar K-L color in Taurus, it would take approximately 
40 magnitudes of visual extinction to produce a K-L color
of 1.5 magnitudes for a naked star and at least 15 magnitudes for a 
Class II PMS star with a disk. Finally, 
there are also 5 objects detected only in L band,
and limits on their K-L color significantly 
exceed 1.5 magnitudes. Altogether we identify 78 candidate protostars from
the list of sources with K-L colors in excess of 1.5 magnitudes.
The positions and JHKL photometry for these 78 candidate protostars are
listed in Table 1.

Figure 6 (a and b) compares the spatial distribution of the 78 protostellar
candidates with that of the cluster membership as a whole. Inspection
of Figure 6a suggests that the 
protostellar candidates are more spatially confined within the
mapped area than are the general cluster members. Figure 6b compares 
the surface density distribution of the protostellar candidates
with that of all other stars in the cluster. The two distributions
are clearly different. The surface density distribution of the general cluster
membership is dominated by a single peak, coincident with the Trapezium stars.
Away from this peak the distribution falls smoothly off with a slight
elongation toward the northwest. In contrast, the surface density
distribution of the protostellar candidates is characterized by 
a more spatially extended and elongated maximum which appears double peaked
and is clearly offset
from the Trapezium stars (and the peak of the cluster surface density
distribution). In addition the overall protostellar surface density distribution
appears more elongated than that of the general cluster membership 
with an orientation in the southwest to northeast direction.
Figure 7 compares the distribution of the protostellar candidates with 
that of millimeter-wave continuum emission from dust in the 
background molecular cloud \citep{jb99}.  The 
protostellar candidate sources have a distribution which is for
the most part, coincident 
with and very similar in extent to that of dust in the well known 
molecular ridge at the back edge of the cluster. This spatial
correspondence with the molecular cloud suggests that 
this group of sources is physically related to the 
molecular cloud and more deeply embedded in it than 
typical cluster members. This is consistent with
the identification of these sources as potential protostars. 

Because protostellar objects can also
be characterized by K-L colors $<$ 1.5 magnitudes and can fall into the 
CTTS region of the color-color diagram, our estimate of the total
protostellar population could be a lower limit. On the other 
hand, it is not clear what fraction of sources located to the right of
the plotted CTTS locus and/or with K-L $>$ 1.5 are true protostars. 
We note that a number of proplyds and a few late type stars
fall to the right of the CTTS locus (see section 4.0 below) and they are not likely
to be protostellar in nature. Whether or not a significant
fraction of the candidate protostars in Table 1 are true
protostellar objects remains to be determined.
Longer wavelength infrared observations with good spatial 
resolution are necessary to enable
a more robust indentification of protostellar objects in this cluster.

\subsection{Stars with Anomalous K-L Colors}

In the JHKL color-color diagram approximately 27 sources (7\% of the
sample) have colors which place them in ``forbidden'' regions either to
the left of the main sequence reddening band or below
and to the right of the CTTS reddening band.
(Figure 8 shows the locations of the various zones the JHKL color-color 
diagram of the Trapezium cluster.)  This is 
almost three times as many sources as the ten sources which
fall into similar regions of
the JHK color-color diagram.  The typical causes for
the presence of stars in these two forbidden regions are large
photometric errors and contamination of photometry due 
to unresolved binaries.
Of the 27 anomalous 
sources, 5 are near our L band detection limit and 
likely characterized by relatively large photometric
L band uncertainties. Most of these sources are in
the forbidden zone to the left of the main sequence
reddening band.  Another 4 objects were found to be
blended stars or sources which were found to be multiple objects
in the Hillenbrand (1997) database but are unresolved in our images.
Thus these two effects account for at least a third of the anomalous sources. 

For our data set another factor 
that could result in a star's placement in one of the 
forbidden regions is  source variability. This is because
the L band observations used in our survey were obtained roughly  
a year after the JHK observations.
To evaluate the possibility of variability as a cause of 
anomalous stellar colors we compared our JHK observations
with those in the compilation of Hillenbrand et al (1998). Of the 256 
sources with JHK magnitudes in both our catalog and the Hillenbrand et
al.  compilation, 25 or 10\% had fluxes which differed in all three 
wavelength bands
by more than 0.2 magnitudes. Of these 12 (or 5\% of the entire
sample of 256 sources) differed
by more than 0.5 magnitudes in all three bands. Thus 
source variability could be responsible for placing 
as many as 5-10\% of the sources in our sample 
in the anomalous regions of the color-color diagram and 
this factor alone could account for all the sources with anomalous colors.
Indeed, 10 of the 
27 sources in the two forbidden zones did vary by more than
0.2 magnitudes in the comparison of our JHK data set with that
of Hillenbrand et al. Finally, it is very unlikely
that variability can account for the vast majority of 
infrared sources located in the excess region (CTTS and
protostellar) of the JHKL color-color diagram since
the expected fraction of variables (5-10\%)
is small compared to the fraction of infrared excess
stars and, moreover, 
already can account for the all sources in the forbidden regions.

\section{Discussion}

Combining L band photometry with JHK observations has enabled us to
determine a more robust estimate of the fraction of stars with 
infrared excess in the Trapezium than has hitherto been possible.
Formally we measure an excess fraction of 80\% $\pm$ 7\%.
Examination of our control fields suggests that contamination 
by background sources with m$_K$ $<$ 12 is insignificant, 
especially given the large extinction provided by the Orion Molecular Cloud.
Thus, the excess fraction we measure corresponds to the excess
fraction of the cluster membership.
If the infrared excesses 
originate in circumstellar disks, then the
vast majority of stars in the Trapezium cluster are presently 
surrounded by and consequently were formed with 
circumstellar disks. Our measurements, therefore,  
appear to confirm earlier suggestions of
a high disk frequency in this young cluster \citep{setal94,hetal98}. 

The assumption of a disk origin for the measured infrared excesses
is bolstered by examination of the observed JHKL colors of 
sources otherwise known to be star/disk systems both in the Taurus
dark clouds and the Trapezium cluster itself. For example, HLL analyzed
observations obtained and compiled by \cite{kh95} of 
known star/disk (class II) sources in the 
Taurus clouds and found JHKL and JHK excess fractions of 100\%
and 69\%, respectively.  We derive very similar statistics
from our JHKL colors of the proplyd population in the Trapezium cluster.
Proplyds are small but resolved nebulous sources
observed in optical HST images
around numerous stars in the Trapezium cluster \citep{owh93,ow96}.
Theoretical considerations indicate that these nebulous objects are most
likely photoevaporating envelopes of circumstellar disks
\citep{jhb98, bsdj98} indicating that the associated stars
have such disks. In our imaging survey we found infrared
sources associated with 112 of 139 proplyds listed by
\citep{ow96} for the area covered by our images. Of these 112,
96 were simultaneously detected in the JHKL bands and 
Figure 9 shows their positions in the JHKL color-color diagram.
We find 97\% of the proplyd stars to have infrared excess.
Furthermore we find that only 70\% of the proplyds exhibit infrared excess 
in the JHK color-color diagram.
Together these results
indicate that JHKL observations are capable of indentifying most, if 
not all, the circumstellar disk sources in a stellar population.
(The close correspondence of JHK and JHKL
excess fractions between class II sources in Taurus and 
proplyds in Orion confirms that source variability has probably not
had a significant effect on our ability to measure excess produced by  
circumstellar disks in the Trapezium cluster.)
The overall distributions of Trapezium excess sources in both
the JHK and JHKL infrared color-color diagrams (Figures 3 \& 4) 
are very similar to the corresponding distributions 
of both the proplyd population in the Trapezium
cluster (Figure 9) and class II sources in the Taurus cloud
(see Figure 8 of HLL). The overall behavior of the Trapezium
excess sources in the JHK and JHKL diagrams is also consistent
with the predictions of circumstellar disk models \citep{mch97}.
Taken together these facts appear to provide compelling evidence
that the infrared excess sources in the Trapezium cluster 
represent a population of circumstellar disk systems and that
JHKL observations provide an accurate census of that population.

As mentioned earlier, there is a population of radio sources
in the cluster whose emission is also thought to arise in the ionized
envelopes of photoevaporated disks \citep{church87,felli93}.  
We detected infrared counterparts for 39 of the 50 VLA sources
within the boundaries
of our imaging survey. Thirty-three were detected in
all four infrared bands. Their locations are shown on the color-color
diagram in Figure 9. Most of them are associated with proplyds, 
independently confirming the identification of the radio 
sources as photoevaporating disks.
Not surprisingly, we find that 87\% of the VLA sources have infrared excesses.
This fraction is somewhat lower than that for the proplyd
population and may indicate that some of the radio sources are
associated with diskless stars. Indeed, 
none of the radio sources without infrared excess
is associated with a proplyd.

Published spectral types are available for
285 stars within the region we imaged \citep{hetal98}. With knowledge
of these spectral types an improved estimate of the disk fraction can
be made for this cluster. The vast
majority (80\%) of these spectroscopically observed 
stars are characterized by spectral types
M3 and earlier.
Since we have set the boundary of the reddening band at the  color of an
M5 star, the excess fraction we have derived for the cluster as a whole
(80\%) from our JHKL observations likely underestimates
the true excess fraction.
Figure 10 shows the color-color diagram for all the
M stars in our imaged region identified by \citet{h97}. Two
different boundaries for the main sequence reddening band are 
drawn at the locations of an M3 and M5 star, respectively. Stars with
spectral types later than M3 are represented as solid circles, while
stars with spectral types between M0 and M3 are represented by open
symbols. The fraction of M0-M3 stars that fall in
the infrared excess region bounded by the (appropriate) M3 reddening band is 
87\% $\pm$ 12\% and the fraction of M4-M6 stars found in the 
reddening band bounded by the M5 boundary is 82\% $\pm$ 17\%. Although
the same within the respective uncertainties, these 
fractions are both formally higher than the 80\% determined
for the whole population measured from the M5 reddening boundary.  
Moreover, the fraction of O-M3 stars that fall into the excess region
as measured from the M3 boundary, is 84\% $\pm$ 8\%. 

Although all the above estimates agree within the uncertainties,
the results suggest that the actual infrared excess/disk fraction
in the Trapezium cluster is close to 85\% and somewhat higher than
our original formal estimate of 80\%.
More significantly, the
high excess fraction is independent of the spectral type or mass of 
stars in the cluster. Table 2 lists the excess/disk fractions
as a function of spectral type (counting from the M5 boundary).
Circumstellar disks appear to form with equally 
high frequency (within the uncertainties)
around all stars with spectral types of F and later, essentially
across the entire span of the stellar
initial mass function down to the hydrogen burning limit.
This is similar to the situation in NGC 2024 where HLL found  
the disk fraction to also be independent of stellar mass.
The earliest type stars in the Trapezium, however, appear to have a lower
disk frequency than the later type stars, although the number of stars in the sample
with spectral types A or earlier is small. 

The overall disk frequency we measure for the
Trapezium is essentially the same
as that (86\%) derived by HLL from similar L band observations of
the NGC 2024 cluster which is also located in
Orion at the same distance from the sun. Both clusters are very young,
although the members of NGC 2024 are more deeply embedded, and
suffer more differential extinction than the Trapezium stars.
Taken together the measurement of similarly high disk
frequencies in the two clusters suggests that disk formation is a
natural consequence of star formation for all stellar masses even
in the environment of a dense stellar cluster. Moreover, 
the timescale for significant inner disk evolution must be longer than
the age of the Trapezium cluster, about 10$^6$ years.
In contrast, JHK observations of NGC 2362 and 
optical spectroscopic studies of the $\lambda$ Ori suggest
very low disk frequencies \citep[$<$ 5\%:][]{allm00,dm99} for those
clusters. This in turn suggests that significant inner disk evolution
does occur on timescales shorter than 4-5 $\times 10^6$ yrs, the estimated 
mean ages of these clusters. Because this important timescale
constrains the duration of planet building within circumstellar
disks, a more robust determination of it is clearly desirable.
We anticipate improved estimates once L band observations of more
clusters in our sample are analyzed. 

There are 78 potential protostars 
in the Trapezium out of some 603 stars detected in both K and L bands
(Table 1). 
The fraction of protostellar (candidate) sources in the Trapezium
is about 13\%, which is similar to the protostellar candidate fraction of
about 14\% derived from L band observations of NGC 2024 by HLL.
For a cluster age of 0.8-1.0 $\times$ 10$^6$ years, the 
lifetime of the protostellar stage would be approximately
10$^5$ years in the Trapezium if all the candidate sources
turned out actually to be protostars. This is typical (within
factors of 2) of protostellar lifetimes
found in Taurus, Ophiuchus, and NGC 2024. Stated another way, if 
a significant fraction (i.e., $\gtrsim$ 50\%) of the protostellar
candidates in the Trapezium
are true protostars, with lifetimes comparable to those
in other regions, then our observations are consistent with a 
relatively constant star formation rate in the cluster up to the 
current epoch. Thus, star formation could be continuing in the molecular 
cloud behind the Orion Nebula a rate similar to that which gave
rise to the foreground cluster!
The molecular ridge behind the Trapezium
is part of a more extended molecular cloud which connects
with the OMC 1, OMC 2, and OMC 3 clouds to the north. 
The extended ridge is also a site of intense protostellar
activity containing at least a dozen Class 0 objects and 
outflow sources \citep{chini97,ybd97,jb99}. This suggests 
a similarly high rate of star formation in the extended ridge,
given that the lifetimes
of Class 0 objects are about an order of magnitude shorter than
those of Class I protostars. Recently \citet{ps99}
have re-analyzed the ages for the stars
in the Orion Nebula Cluster and have suggested that the rate of
star formation in this region has been accelerating with time with the
most vigorous episode taking place in the last two million years.
They have also suggested that the formation of O stars in the cluster
should squelch further star formation with the result that the 
star formation rate should be drastically decreased at present or
in the near-future.  Palla and Stahler estimate that the timescale
for this steep decline should be less than 10$^6$ years. If the 
above estimate of the protostellar population is close to the
true size, then it appears that the predicted rapid decline has
not yet occured (primarily because the O stars have not completely
cleared the region of dense star forming gas). However, our identifications
of protostars in the Trapezium region 
are not secure, and longer wavelength observations
are necessary to obtain a more accurate estimate of the protostellar
population in the cluster and a more robust determination of the 
current star formation rate in the cloud behind it.

\section{Summary}

We have obtained a sensitive L band (3.4 \micron) survey of 
a $\sim$ 36 arcmin$^2$ region centered on the Trapezium cluster in
Orion. We have used these observations in conjunction with new JHK imaging 
observations of the same region to produce 
a relatively  unambiguous census of the circumstellar disk
population in the cluster and to compile a list of candidate protostars
within the region. The primary results and conclusions derived from 
our study can be summarized as follows:

\noindent
1.\ We detected 391 stars in the surveyed region
with K and L $<$ 12.0 magnitudes. From analysis of the JHKL
color-color diagram we find that 80\% $\pm$ 7\% of the stars exhibit 
detectable excess emission and are likely surrounded by circumstellar
disks.  
A subsample of 285 of these stars for which there are 
published spectral types, has a slightly higher 
excess/disk fraction of 85\%. More significantly,
within our statistical uncertainties, this high excess/disk fraction
is independent of spectral type for stars with spectral types 
between F-M. Thus the probability of disk
formation around a star at the time of its birth
is both extremely high and independent of stellar mass across 
essentially the entire stellar mass spectrum down to the 
hydrogen burning limit. The disk fraction (42\%) for the higher mass O, 
B and A stars is lower however, and may suggest either a lower
probability for disk formation or more rapid disk dispersal times
for these stars. Overall the results presented in this paper 
confirm and improve upon previous studies which also suggested a
high disk fraction in the cluster. 
 
\noindent
2.\ Of known proplyds in our surveyed region, 97\% have
infrared  L band excess.
The fraction of radio continuum emitting stellar
systems with L band excess is found to be 87\%. These results 
provide additional confirmation for our identification of infrared excess
sources with circumstellar disks, and they illustrate 
the effectiveness of L band observations
for obtaining an accurate census of star/disk systems in the cluster. 

\noindent
3.\ The disk fraction in the Trapezium cluster is very similar
to that (86\% $\pm$ 8\%) recently obtained from a sensitive  L band
survey of NGC 2024, the other rich and very young embedded cluster in
Orion (Haisch, Lada \& Lada 2000). Thus our results confirm the findings
of the NGC 2024 study, which suggested that, independent of mass, 
stellar birth in the environment of a rich and dense 
cluster does not diminish the probability of disk formation 
around a new star. Indeed,
in both these clusters, the vast majority of stars, spanning 
essentially the entire range of the stellar mass function,
were
born with circumstellar disks and consequently the potential
to form planetary systems. 
Moreover, the high disk frequencies found in these two clusters
indicate that disk lifetimes are on the order of the age of the clusters, 
$\sim10^6$ years, or longer.

\noindent
4.\  Seventy-eight of the sources in our survey possess K-L colors
typical of deeply embedded (A$_V \gtrsim
20-40$ magnitudes) young stellar objects. Only about half of these
stars are detectable at the J and H bands, but most of these
are characterized by extreme infrared excess falling outside
the CTTS region (see Figure 3 \& 4) of the JHKL color-color diagram.
The spatial distribution of these very red stars is very similar to
that of the dense molecular ridge behind the cluster.
These properties are characteristic of protostellar objects, and we
identify these objects as candidate protostars (Class I objects). If
the size of the protostellar population is even a modest fraction of our
candidate protostar sample, then star formation may be continuing within
the molecular ridge at a rate similar to that which produced the 
stars in the foreground Trapezium cluster. Sensitive, longer wavelength
observations with good spatial resolution are necessary to determine
the fraction of L band candidates that are truly protostellar in
nature and evaluate the present level of star forming activity in the
Trapezium region.

\acknowledgements
\noindent
We thank Doug Johnstone for providing us with the 850 $\mu$m map 
of the OMC 1 region.
EAL acknowledges support from a Research Corporation Innovation
Award and a Presidential Early Career Award 
for Scientists and Engineers (NSF AST 9733367).
AM is a Smithsonian Predoctoral Fellow.
K.E.H. gratefully acknowledges support from a 
NASA Florida Space Grant Fellowship and an ISO grant through JPL \#96104.
We also acknowledge support from an ADP (WIRE) grant NAG 5-6751.

\clearpage


\clearpage

\begin{deluxetable}{ccccccc}
\small
\tablecaption{Candidate Protostellar Sources.\label{table:proto}}
\tablewidth{0pt}
\tablehead{
\colhead{ID}&\colhead{RA (J2000)}&\colhead{DEC (J2000)} &
\colhead{ L }&
\colhead{ J-H }&\colhead{ H-K }&\colhead{ K-L }
}
\startdata
\cutinhead{Candidates detected at only 3.5$\mu$m\tablenotemark{a}}
   1 &  5 35 14.440 &  -5 23 51.54 &  7.91 & -- & -- & $>$ 6.09 \\ 
   2 &  5 35 14.540 &  -5 23 55.23 &  9.10 & -- & -- & $>$ 4.90\\ 
   3 &  5 35 13.410 &  -5 23 29.77 &  9.77 & -- & -- &$>$ 4.23 \\ 
   4 &  5 35 16.630 &  -5 21 52.86 & 10.93 & -- & -- &$>$ 3.08\\ 
   5 &  5 35 17.830 &  -5 24 47.58 & 11.14 & -- & -- &$>$ 2.87\\ 
\cutinhead{JHKL Candidates: K-L $>$ 1.5; J-H $>$ 1.0}
   6 &  5 35 18.830 &  -5 22 23.20 &  9.51 &  1.55 &  0.79 &  1.51 \\ 
   7 &  5 35 14.310 &  -5 23 8.500 &  8.77 &  1.69 &  0.91 &  1.51 \\ 
   8 &  5 35 18.860 &  -5 20 31.21 &  9.74 &  1.08 &  0.67 &  1.52 \\ 
   9 &  5 35 10.910 &  -5 22 46.60 &  9.24 &  1.37 &  1.01 &  1.52 \\ 
  10 &  5 35 15.870 &  -5 22 33.03 &  9.77 &  1.08 &  1.08 &  1.53 \\ 
  11 &  5 35 20.990 &  -5 20 43.33 & 10.68 &  2.66 &  1.54 &  1.56 \\ 
  12 &  5 35 15.390 &  -5 21 39.82 & 10.68 &  2.17 &  1.20 &  1.58 \\ 
  13 &  5 35 5.6900 &  -5 23 45.19 & 10.59 &  1.61 &  1.22 &  1.58 \\ 
  14 &  5 35 5.4000 &  -5 24 10.43 &  8.52 &  1.14 &  0.67 &  1.58 \\ 
  15 &  5 35 14.940 &  -5 21 1.150 & 11.15 &  1.37 &  1.07 &  1.59 \\ 
  16 &  5 35 13.550 &  -5 23 59.77 &  9.71 &  2.21 &  1.62 &  1.60 \\ 
  17 &  5 35 12.860 &  -5 21 34.22 &  8.32 &  3.41 &  1.98 &  1.62 \\ 
  18 &  5 35 13.610 &  -5 19 55.32 &  6.67 &  1.65 &  1.08 &  1.65 \\ 
  19 &  5 35 17.540 &  -5 21 45.82 &  7.86 &  1.14 &  0.47 &  1.66 \\ 
  20 &  5 35 19.810 &  -5 22 21.80 &  8.75 &  1.16 &  0.95 &  1.67 \\ 
  21 &  5 35 18.660 &  -5 23 14.12 &  5.83 &  1.74 &  1.59 &  1.67 \\ 
  22 &  5 35 17.910 &  -5 20 55.75 &  9.77 &  1.16 &  1.04 &  1.71 \\ 
  23 &  5 35 16.280 &  -5 22 10.68 &  8.62 &  1.29 &  1.02 &  1.71 \\ 
  24 &  5 35 7.7300 &  -5 21 1.940 &  8.85 &  2.06 &  1.32 &  1.75 \\ 
  25 &  5 35 22.260 &  -5 21 42.68 & 11.43 &  1.12 &  1.09 &  1.75 \\ 
  26 &  5 35 16.380 &  -5 24 37.23 &  9.68 &  1.75 &  1.30 &  1.78 \\ 
  27 &  5 35 19.530 &  -5 21 49.34 &  9.75 &  2.52 &  2.02 &  1.79 \\ 
  28 &  5 35 18.230 &  -5 24 13.37 & 11.49 &  1.07 &  1.00 &  1.81 \\ 
  29 &  5 35 13.750 &  -5 22 22.23 &  7.30 &  1.11 &  1.16 &  1.82 \\ 
  30 &  5 35 17.760 &  -5 23 42.62 & 10.40 &  1.23 &  0.83 &  1.85 \\ 
  31 &  5 35 12.200 &  -5 24 56.38 &  9.09 &  3.91 &  1.83 &  1.86 \\ 
  32 &  5 35 11.490 &  -5 23 51.99 & 10.49 &  2.12 &  1.19 &  1.89 \\ 
  33 &  5 35 27.740 &  -5 21 18.70 &  9.71 &  3.04 &  1.83 &  1.90 \\ 
  34 &  5 35 24.600 &  -5 21 4.310 &  8.95 &  3.46 &  2.12 &  1.93 \\ 
  35 &  5 35 13.710 &  -5 21 36.22 &  8.45 &  3.83 &  2.36 &  1.96 \\ 
  36 &  5 35 11.200 &  -5 22 38.01 &  7.71 &  2.62 &  1.72 &  1.97 \\ 
  37 &  5 35 19.580 &  -5 20 1.980 &  9.07 &  2.72 &  1.68 &  1.98 \\ 
  38 &  5 35 20.250 &  -5 25 4.200 &  9.30 &  1.51 &  1.29 &  2.00 \\ 
  39 &  5 35 11.520 &  -5 25 53.28 & 10.92 &  1.85 &  1.67 &  2.03 \\ 
  40 &  5 35 13.790 &  -5 21 59.88 &  4.55 &  2.94 &  2.16 &  2.55 \\ 
  41 &  5 35 19.380 &  -5 25 42.52 &  7.43 &  3.05 &  2.66 &  2.81 \\ 
  42 &  5 35 16.080 &  -5 22 54.42 &  7.80 &  1.57 &  1.87 &  2.85 \\ 
  43 &  5 35 14.350 &  -5 22 32.95 &  5.72 &  2.81 &  2.71 &  2.99 \\ 
\cutinhead{KL Candidates: No J or H, K-L $>$ 1.5}
  44 &  5 35 21.790 &  -5 20 7.910 & 11.92 & -- &  1.64 &  1.52 \\ 
  45 &  5 35 18.530 &  -5 21 50.28 & 11.10 & -- &  1.76 &  1.54 \\ 
  46 &  5 35 13.580 &  -5 23 55.45 &  8.38 & -- &  2.29 &  1.67 \\ 
  47 &  5 35 21.240 &  -5 23 17.04 & 11.18 & -- &  1.56 &  1.70 \\ 
  48 &  5 35 10.400 &  -5 22 59.92 & 10.90 & -- &  2.20 &  1.76 \\ 
  49 &  5 35 15.680 &  -5 23 39.63 &  8.58 & -- &  1.57 &  1.79 \\ 
  50 &  5 35 12.490 &  -5 24 38.08 & 11.26 & -- &  2.41 &  1.81 \\ 
  51 &  5 35 19.370 &  -5 23 6.650 & 10.91 & -- &  1.39 &  1.81 \\ 
  52 &  5 35 17.870 &  -5 22 3.110 &  8.94 & -- &  2.03 &  1.85 \\ 
  53 &  5 35 27.180 &  -5 23 15.94 & 11.84 & -- &  1.33 &  1.88 \\ 
  54 &  5 35 19.070 &  -5 23 7.360 & 10.40 & -- &  1.29 &  1.89 \\ 
  55 &  5 35 10.710 &  -5 22 20.62 & 11.06 & -- &  2.32 &  1.91 \\ 
  56 &  5 35 18.120 &  -5 20 29.80 & 10.87 & -- &  2.33 &  1.95 \\ 
  57 &  5 35 10.710 &  -5 23 32.91 & 11.78 & -- &  2.12 &  1.97 \\ 
  58 &  5 35 19.940 &  -5 21 54.10 &  9.63 & -- &  1.78 &  1.99 \\ 
  59 &  5 35 15.290 &  -5 21 29.07 & 10.57 & -- &  2.05 &  1.99 \\ 
  60 &  5 35 20.010 &  -5 25 22.31 & 11.81 & -- & -- &  2.00 \\ 
  61 &  5 35 23.870 &  -5 23 33.25 & 10.53 & -- &  0.70 &  2.03 \\ 
  62 &  5 35 17.880 &  -5 21 53.65 &  9.76 & -- &  1.84 &  2.06 \\ 
  63 &  5 35 10.960 &  -5 21 46.75 & 10.74 & -- &  2.50 &  2.06 \\ 
  64 &  5 35 11.740 &  -5 23 33.67 & 11.38 & -- & -- &  2.07 \\ 
  65 &  5 35 11.840 &  -5 21 0.700 &  7.95 & -- &  2.55 &  2.13 \\ 
  66 &  5 35 17.540 &  -5 22 0.640 & 11.35 & -- & -- &  2.15 \\ 
  67 &  5 35 23.900 &  -5 20 10.36 & 11.26 & -- &  2.55 &  2.17 \\ 
  68 &  5 35 14.540 &  -5 23 3.800 & 10.08 & -- &  1.03 &  2.40 \\ 
  69 &  5 35 17.360 &  -5 22 45.94 &  9.78 & -- &  1.60 &  2.47 \\ 
  70 &  5 35 13.360 &  -5 20 51.96 & 11.46 & -- & -- &  2.51 \\ 
  71 &  5 35 24.580 &  -5 19 55.01 & 11.14 & -- &  3.23 &  2.53 \\ 
  72 &  5 35 5.1700 &  -5 25 1.080 & 12.15 & -- & -- &  2.53 \\ 
  73 &  5 35 20.280 &  -5 19 58.09 &  8.95 & -- &  2.60 &  2.55 \\ 
  74 &  5 35 11.320 &  -5 24 38.23 & 10.10 & -- &  3.28 &  2.56 \\ 
  75 &  5 35 14.280 &  -5 23 4.330 & 10.05 & -- & -- &  2.73 \\ 
  76 &  5 35 15.200 &  -5 22 37.01 &  8.85 & -- &  2.92 &  2.76 \\ 
  77 &  5 35 14.730 &  -5 22 29.87 &  7.70 & -- &  3.18 &  2.81 \\ 
  78 &  5 35 13.820 &  -5 23 40.15 &  6.99 & -- & -- &  4.85 \\ 

\enddata

\tablenotetext{a}{K-L upper limits calculated assuming K $>$ 14.0
magnitude.}
\end{deluxetable}

\clearpage


\clearpage
 
\begin{deluxetable}{lrrcc}
\small
\tablecaption{IR Excess Fraction vs Spectral Type \label{table:spt_table}}
\tablewidth{0pt}
\tablehead{
\colhead{Spectral Type(s)\tablenotemark{1}}&
\colhead{$N_{Region}$\tablenotemark{2}}&
\colhead{$N_{Detect}$\tablenotemark{3}}&
\colhead{$JHK_{Excess}$ (\%)}&
\colhead{$JHKL_{Excess}$ (\%)\tablenotemark{4}}
}
\startdata
OBA     &  12 &  12 &   8   & 42    \\
FG      &  10 &   9 &  44   & 78    \\
K       &  91 &  87 &  44   & 82    \\
M       & 191 & 177 &  45   & 81    \\
M0 - M3 & 129 & 121 &  46   & 80    \\
M4 - M6 &  63 &  57 &  41   & 82    \\
All     & 304 & 285 &  45   & 78    \\
\enddata
\tablenotetext{1}{ Spectral types taken from Hillenbrand (1997).}
\tablenotetext{2}{ Number of stars with Spectral Types within survey boundaries.} 
\tablenotetext{3}{ Number of stars with Spetcral Types and JHKL Photometry. }
\tablenotetext{4}{ Counting from the M5 boundary of the reddening band and 
using the Cohen et al. (1981) IR reddening law.}
\end{deluxetable}
 
\clearpage

\clearpage

\centerline{\bf Figure Captions}

\figcaption[trap_images.eps]
{a) A false color JHK image of the Trapezium region
within the Great Orion Nebula. K band (2.2 \micron) emission is coded as red,
H band (1.65 \micron) emission as green and J band (1.25 \micron)
emission as blue. b) The L band (3.5 micron) mosaic image of
the Trapezium cluster
Both images were constructed from the $3\times3$ mapping mosaic described in
the text and have resolutions of 0.6\arcsec/pixel
with seeing of 1.2 - 1.6\arcsec\  FWHM.}


\figcaption[comet.eps]{a) Close up L band image of the extended
comet or jet-like feature found northeast of $\theta^1C$ Ori and under
the ``Dark Bay'' prominent in optical images of the Trapezium cluster. 
Also plotted are contours of relative surface brightness which are spaced
at intervals of 0.025 dex.
 b) J band image of the same field. The cometary feature does not
appear on the J band image although a dark lane
which is roughly orthogonal to it is very prominent.            
Relative surface brightness contours of the dark lane are plotted
spaced and in intervals of 0.15 dex.
\label{comet}}

\figcaption[trap_jhk.eps]
{JHK color-color diagram for the Trapezium cluster.
The locus of points representing the intrinsic infrared colors
of main sequence stars is represented by a solid curved line. The 
locus of intrinsic colors of giant stars is represented as a dotted line
which partially overlaps with the main sequence locus.
The locus of unreddened Classical T Tauri star colors \citep{mch97} is
represented by a solid straight line. Dot-dashed lines
represent the boundaries of the two reddening bands and are parallel
to the reddening vector. The length of the arrow, also drawn parallel to the reddening
vector, corresponds to the displacement 
produced by 10 magnitudes of {\it visual} extinction.
Stars which fall to the right of the main sequence reddening band
possess infrared excess. This diagram plots the positions of all
stars in our survey with K and L magnitudes 
$<$ 12.0 and with L band photometric uncertainties $<$ 0.25. 
\label{trap_jhk}}

\figcaption[trap_jhkl.eps]{JHKL color-color diagram of all Sources with K = L
$<$ 12.0 and with L errors $<$ 0.25. Otherwise same as figure 3. 
\label{trap_jhkl}}

\figcaption[trap_kl_hist.eps]{Frequency distribution of all sources
in our survey detected in both the K and L bands.
Those sources lacking either J or H band photometry are indicated. 
\label{trap_kl_hist}}

\figcaption[trap_spatial.eps] 
{a) Spatial distributions of cluster sources (open circles) and candidate 
protostars (filled circles). The protostellar candidates appear to be more
spatially confined than the main cluster population.
Large white stars label the locations of bright stars with spectral types
B3 and earlier.  Large black stars mark the positions of sources
detected only at L band.
b) Surface density distributions of cluster sources (grey scale) and
candidate protostars (contours). The two distributions are clearly
different. These maps were constructed from star counts obtained by 
sampling the source distributions at the Nyquist interval
with a square box 0.1 pc (51.6 arc sec) in size. For the grey scale
plot, the levels are spaced at intervals of 500 stars pc$^{-2}$
starting from a level of 750 stars pc$^{-2}$. For the contour map
the contours are spaced at intervals of 100 stars pc$^{-2}$ starting
from a level of 350 stars pc$^{-2}$. The peaks in the two distributions
reach values of 4250 and 750 stars pc$^{-2}$, respectively.
\label{trap_spatial}}

\figcaption[scuba_proto.eps]{Comparison of the distribution of
protostellar candidate sources with 
the $850\mu\mbox{m}$ SCUBA map from Johnstone and Bally (1999).
Symbols are the same as in Figure \ref{trap_spatial}.
In addition, the CS $\mbox{J} = 1 \rightarrow 2$
continuum sources from Mundy et al (1986) are plotted with triangles
and the locations of the IRC2 and
BN-KL proto-stellar sources \citep{dlrcp93}
are shown with crosses.
The north-south distribution 
of the candidate proto-stellar objects closely follows the dust continuum map which
traces the
location of the dense star forming gas in the OMC-1 cloud complex.
\label{fig:scuba_proto}
}

\figcaption[]{JHKL color-color diagram of the cluster with the positions
of sources with anomalous colors indicated
by filled circles.
The various zones of the diagram discussed in the text are also shown.}

\figcaption[trap_prop_rad_jhkl.eps]{JHKL color-color diagram for all detected
proplyd \citep{ow96} and radio sources \citep{felli93}. Examination
of this diagram shows that 97\% of the optical proplyds exhibit infrared
excess emission. This excess emission originates in circumstellar disks
around the proplyd stars and this plot illustrates the effectiveness of 
using the JHKL color-color diagram to identify circumstellar disk
sources. 
Filled circles inside of open circles represent radio sources identified
as proplyds. Examination of the figure indicates that 87\% of the radio
emitting stars possess infrared excess which originates in
circumstellar disks.
\label{trap_prop_rad_jhkl}}

\figcaption[trap_mstars_jhkl.eps]{The locations of
M stars in the JHKL color-color diagram. Spectral types were taken
from Hillenbrand (1997). A total of 
121 of 129 M0-M3 stars and 56 of 62 M4-M6 stars
were detected at all JHK and L wavelengths. Plotted are four reddening
vector boundaries set by the colors of giants , M3/A0 dwarfs,
M5 dwarfs and the tip of the CTTS locus, respectively.
\label{trap_mstars_jhkl}}

\epsscale{1.0}
 
%
 
\clearpage
\plotone{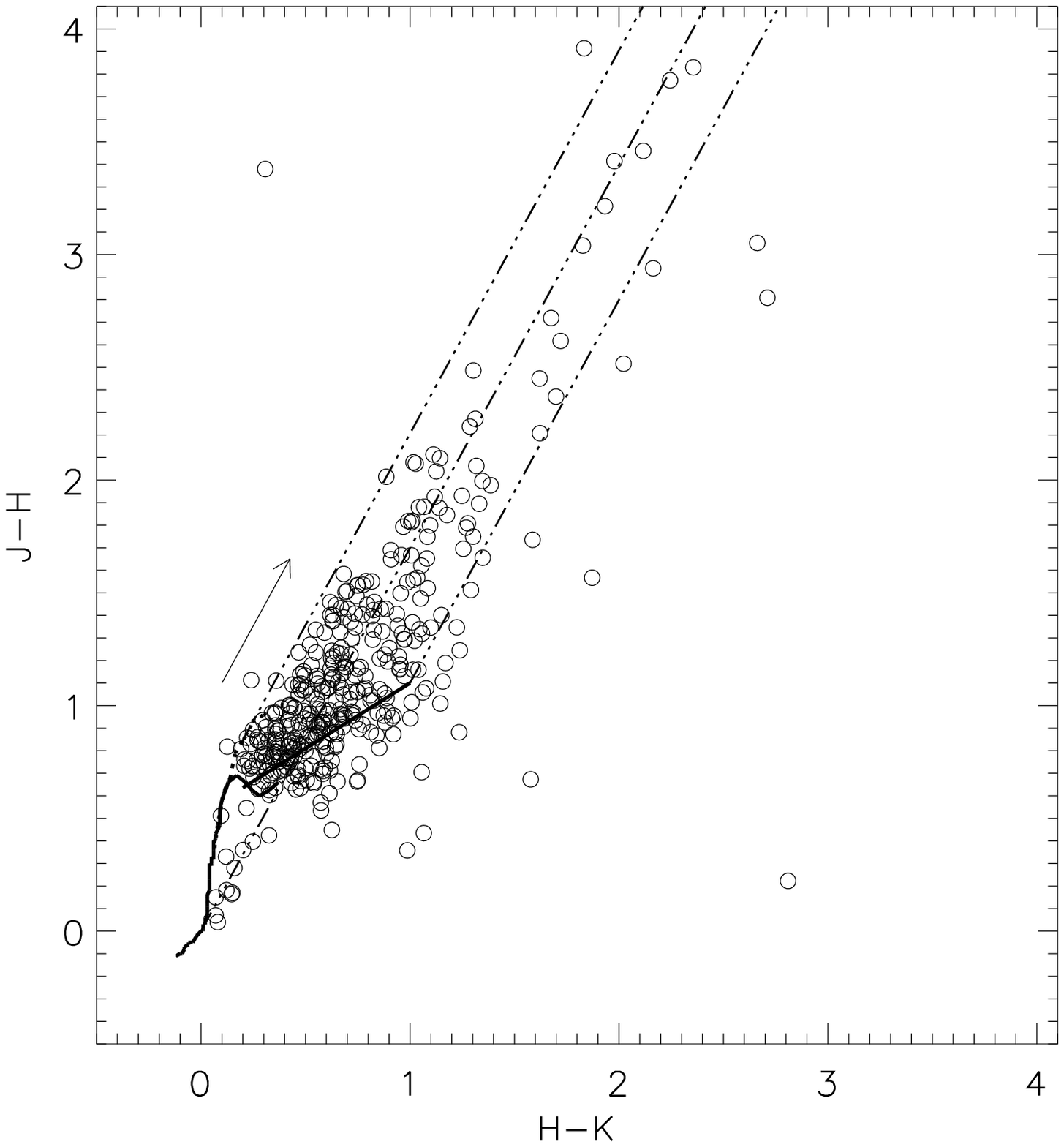}
 
\clearpage
\plotone{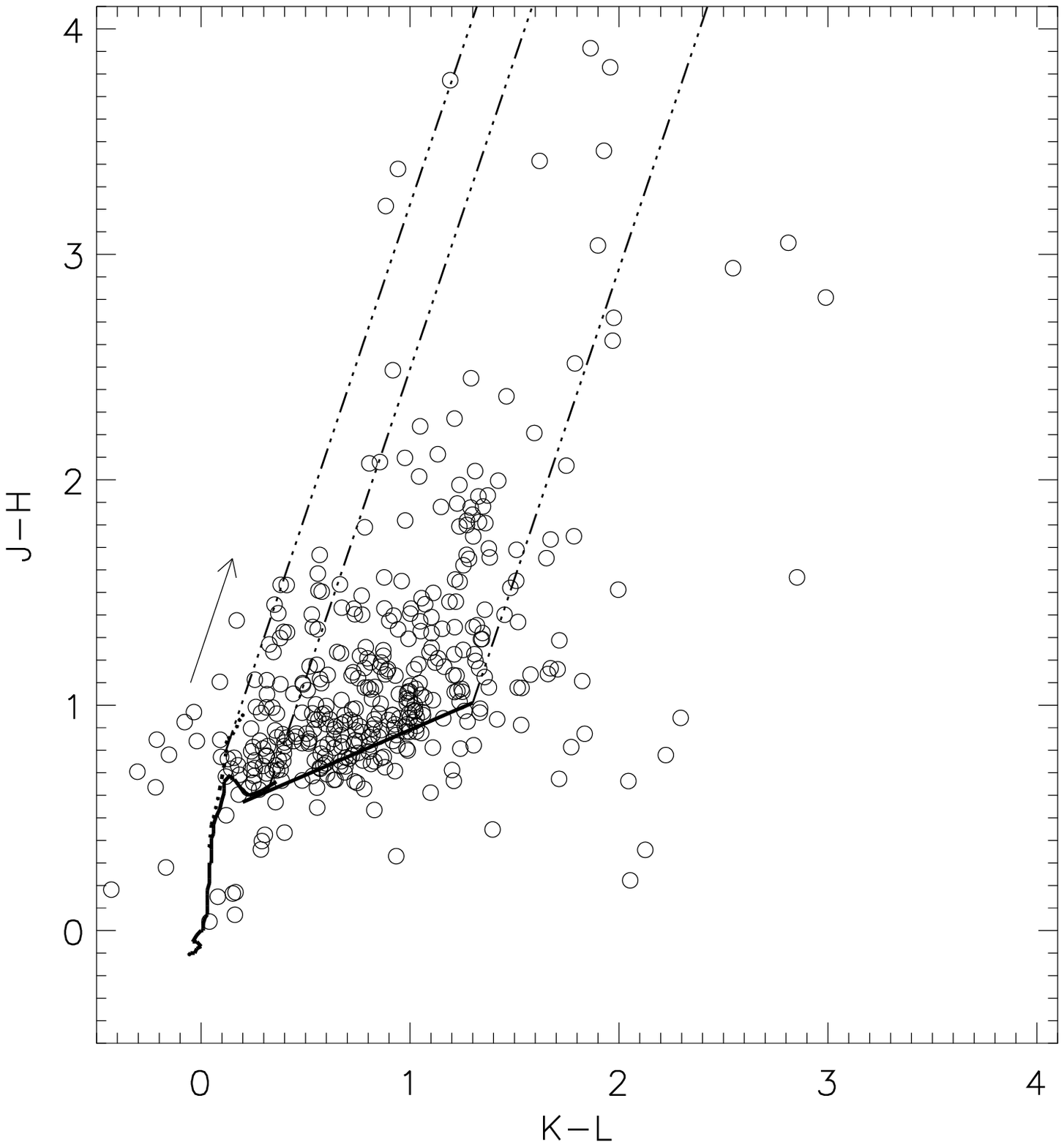}
 
\clearpage
\plotone{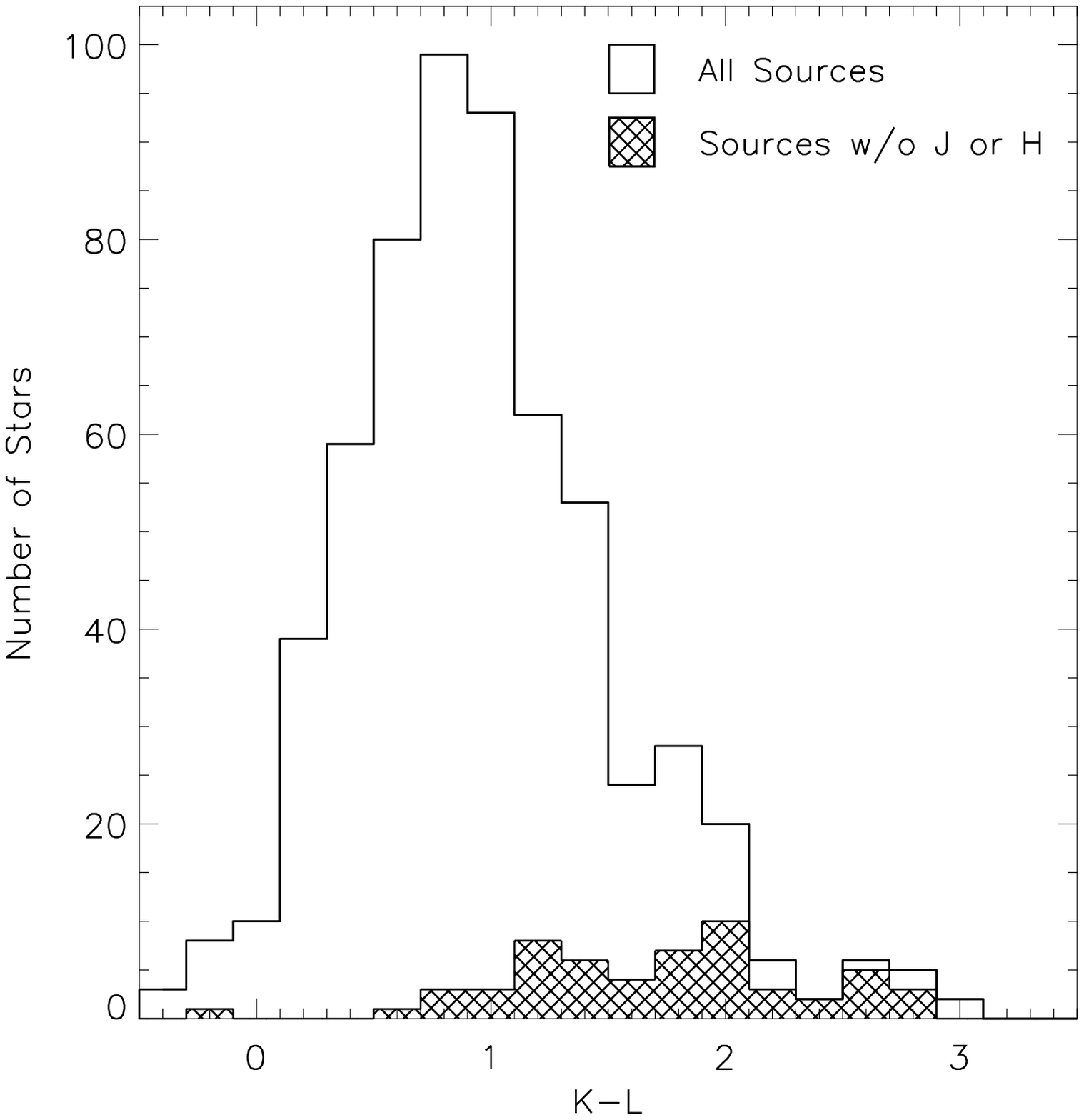}
 
\clearpage
\epsscale{0.5}
\plotone{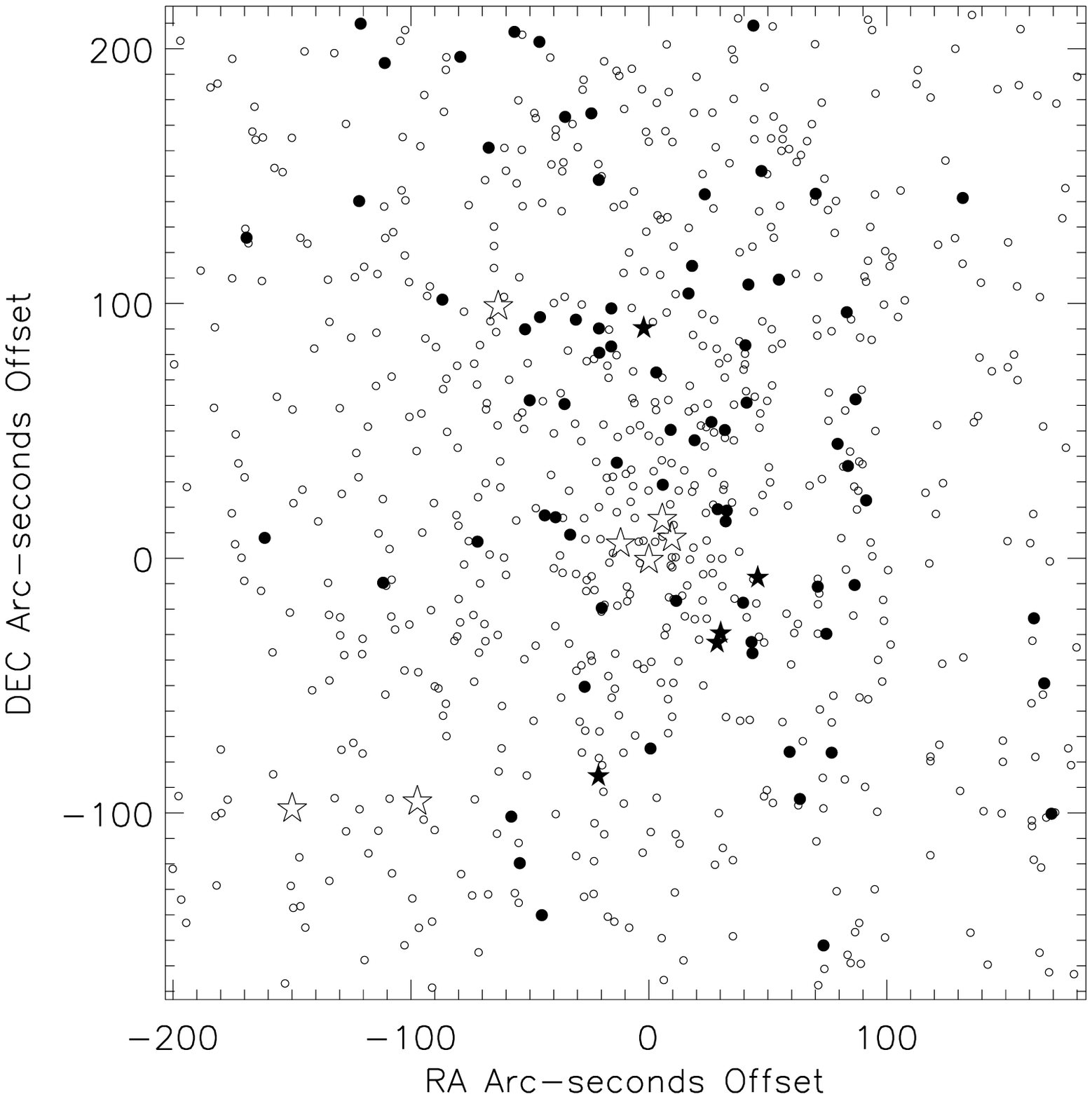}
\\[0.25in]
\plotone{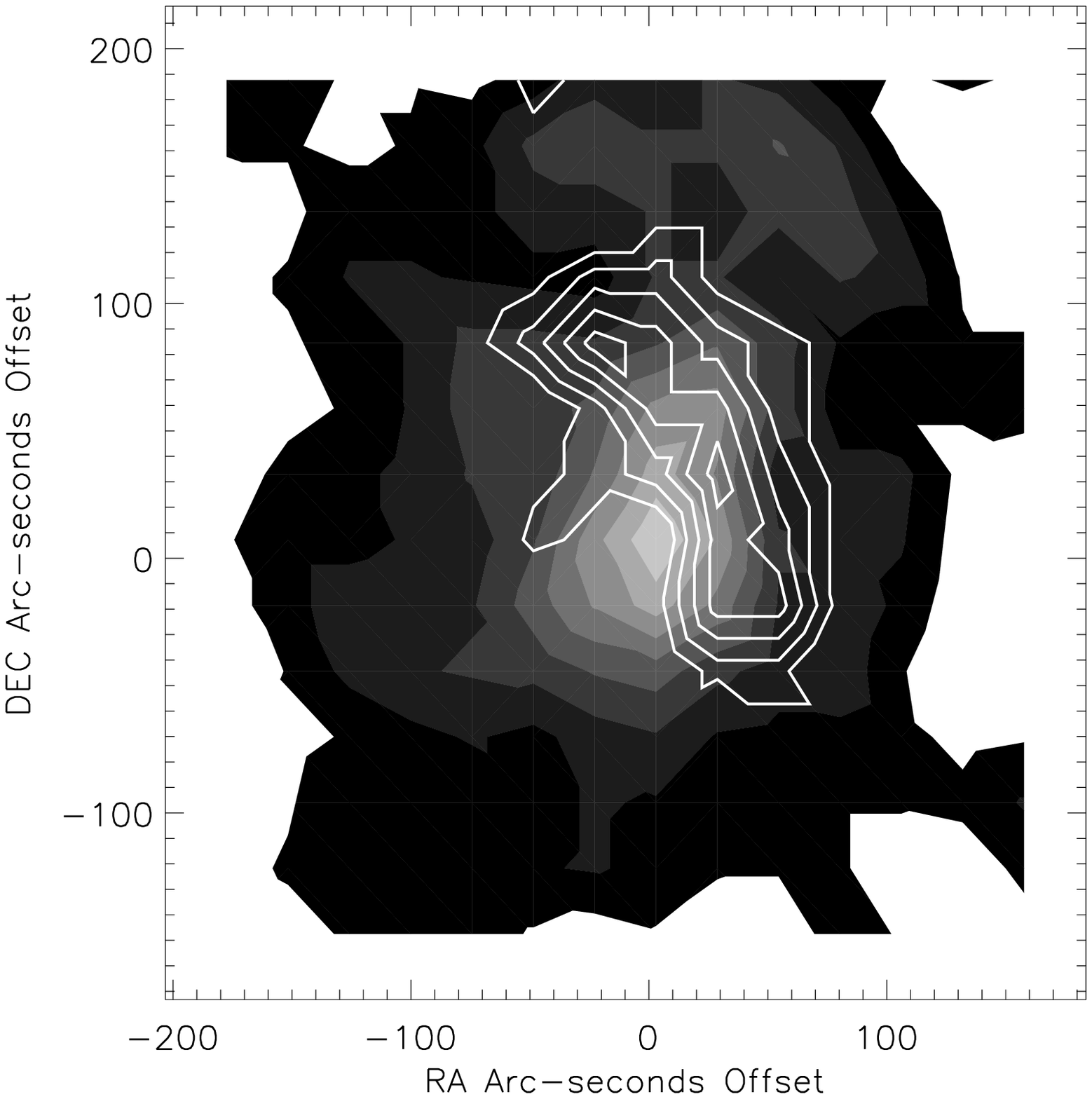}
\epsscale{1.00}

\clearpage
\plotone{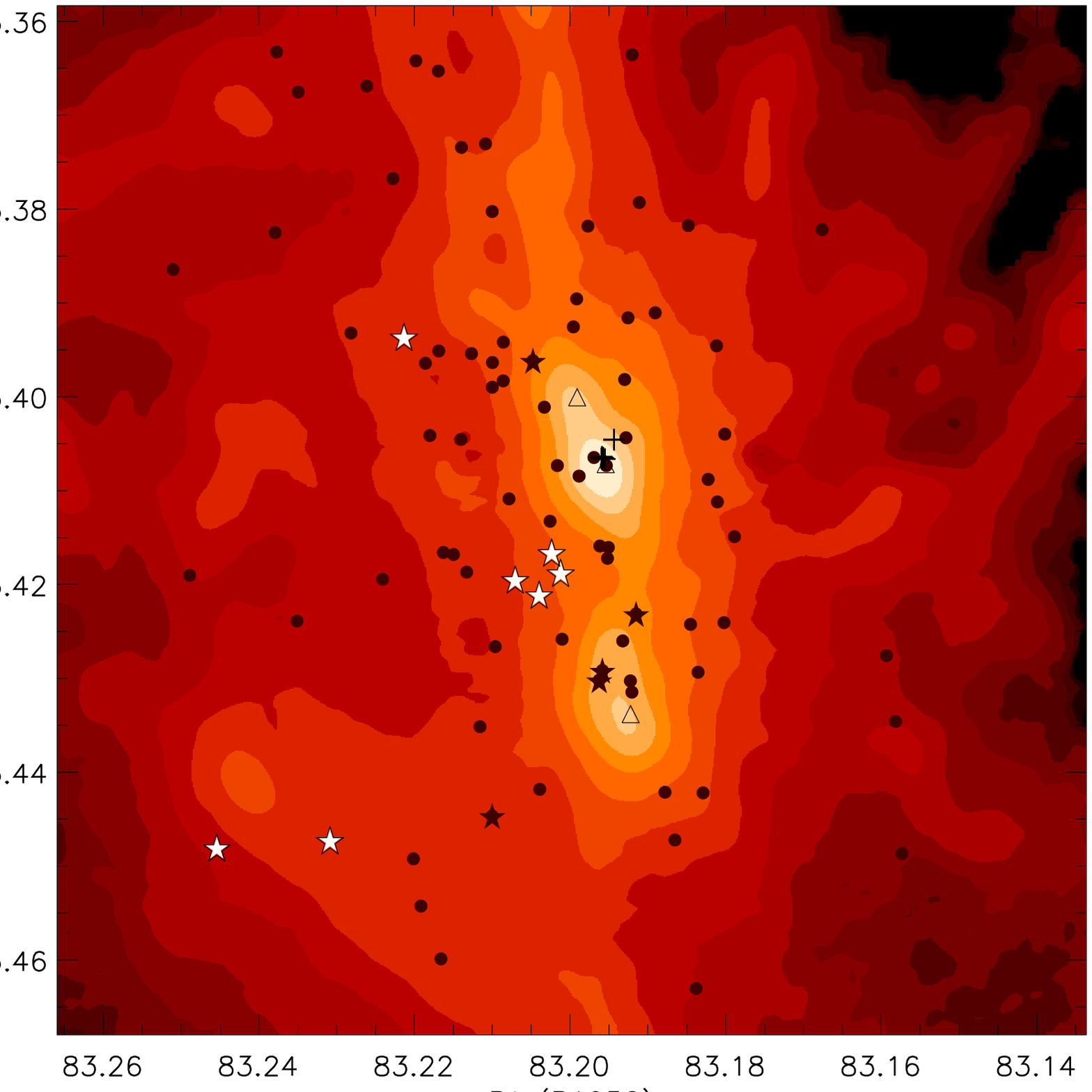}
 
\clearpage
\plotone{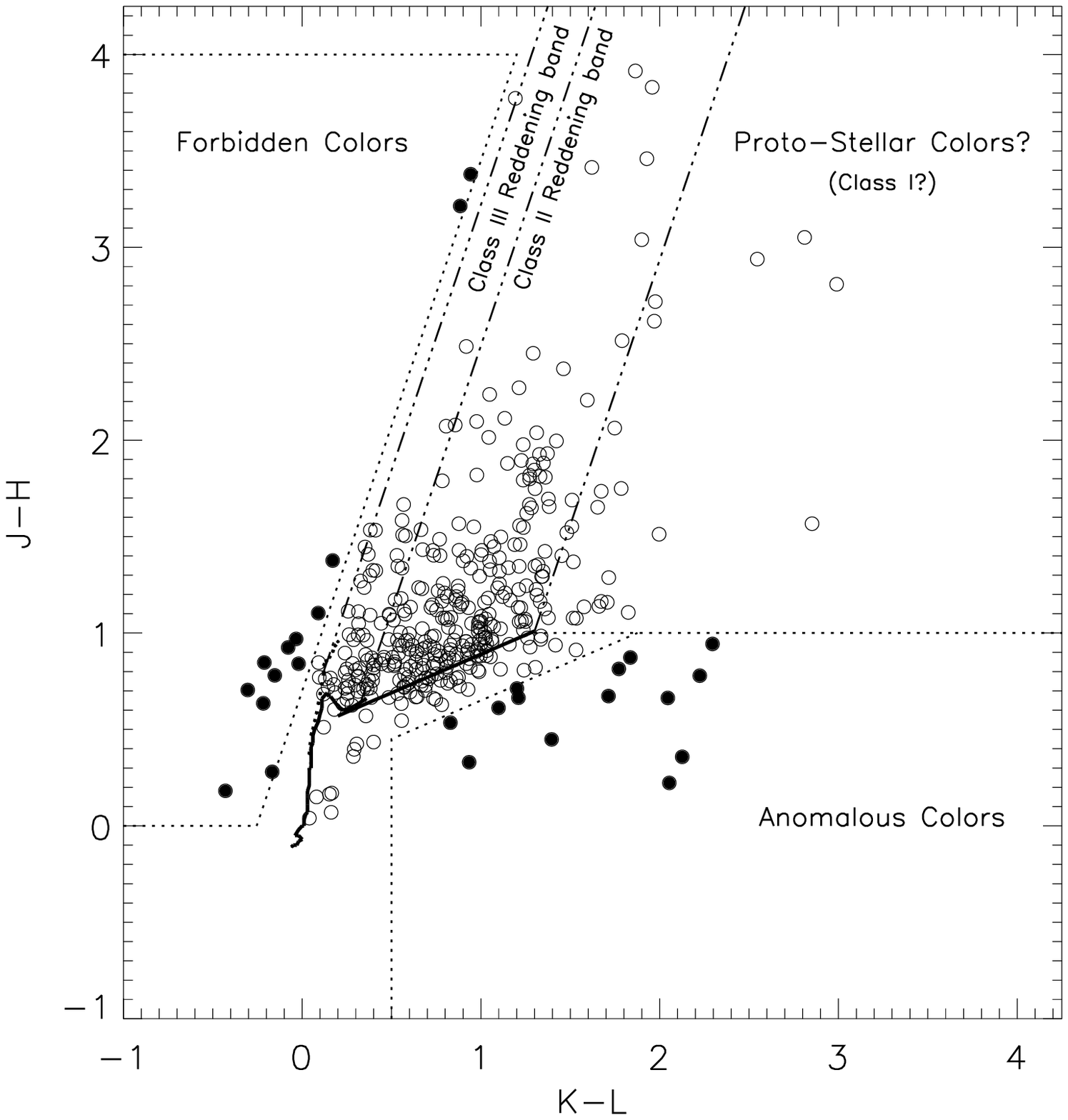}
 
\clearpage
\plotone{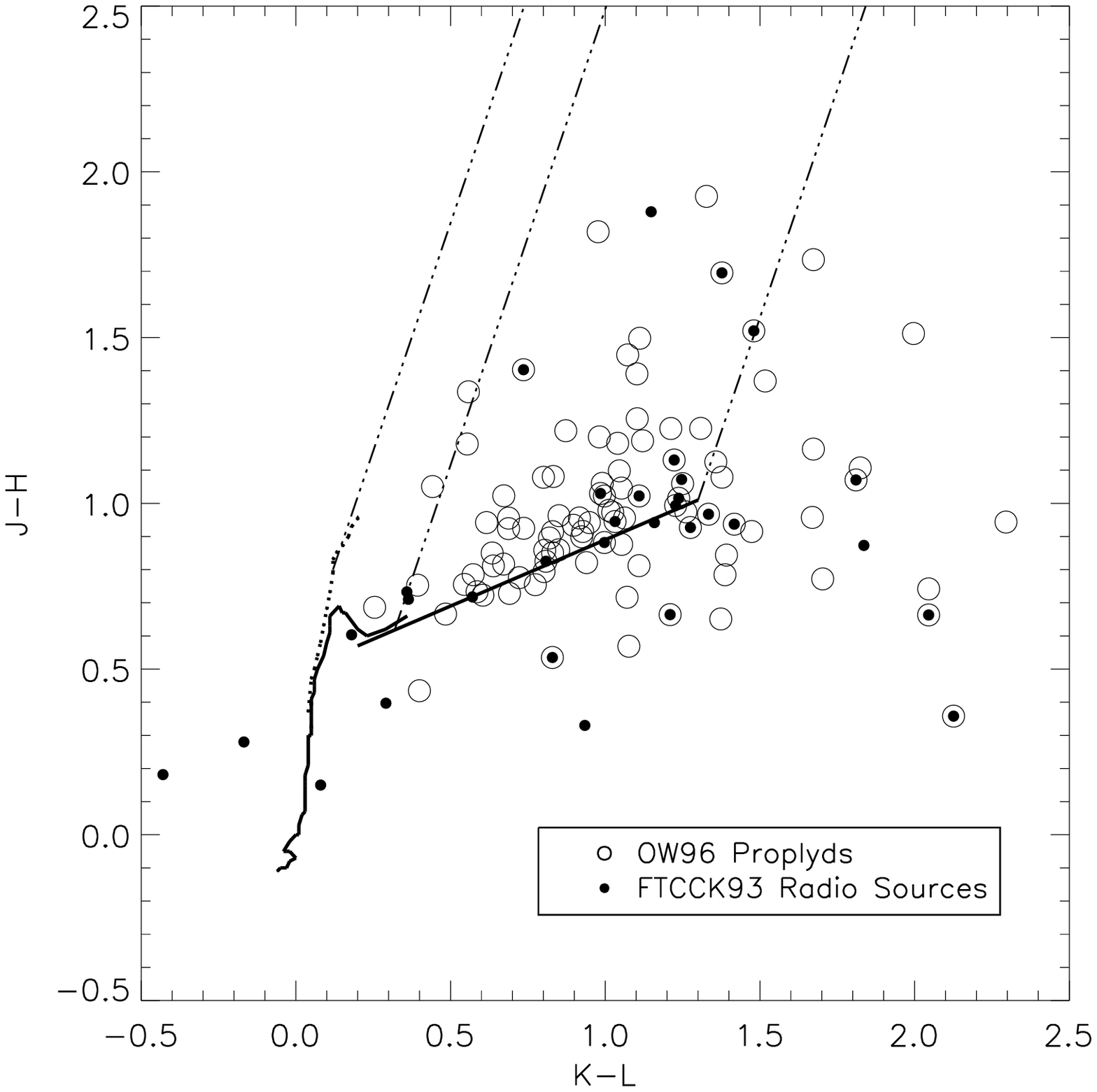}
 
\clearpage
\plotone{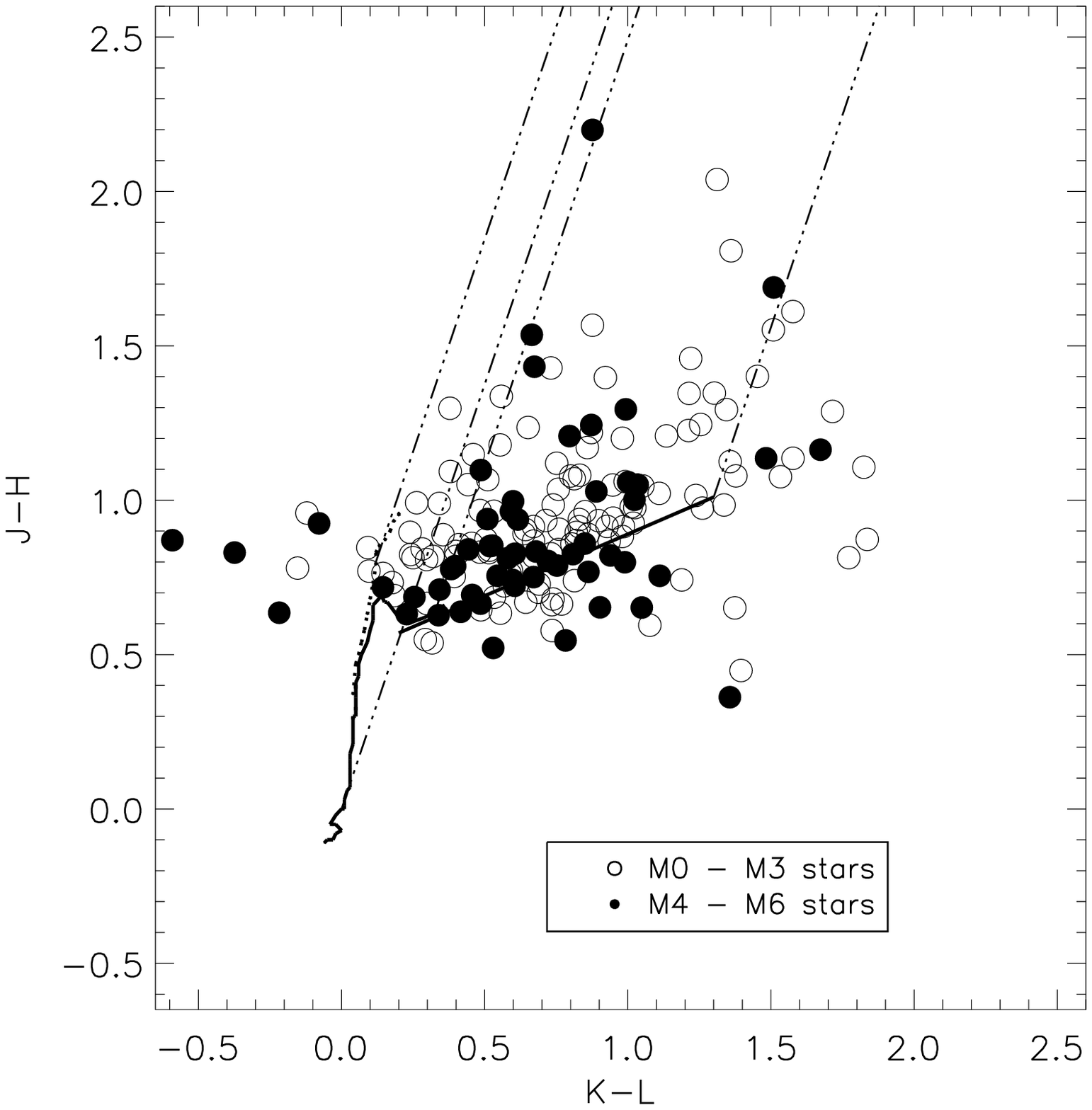}
 
\clearpage


\begin{thebibliography}{}

\bibitem[Adams, Lada, \& Shu(1987)]{als87} Adams, F.C., Lada, C.J.
\& Shu, F.H. 1987, \apj, 312, 788.

\bibitem[Ali \& DePoy(1995)]{ad95} Ali, B. \& DePoy, D.L. 1995, \aj, 110, 2415.

\bibitem[Alves et al.(2000)]{allm00} Alves, J., Lada, C.J., Lada, E.A.,
\& Muench, A.A. 2000 in preparation.

\bibitem[Bally et al.(1998)]{bsdj98} Bally, J., Sutherland, R.S.,
Devine, D. \& Johnstone, D. 1998 \aj, 116, 293.

\bibitem[Beckwith(1999)]{svwb99} Beckwith, S.V.W. 1999, in
The Origin of Stars and Planetary Systems, eds. C.J. Lada \& N.D.
Kylafis, Kluwer: Dordrecht, p. 579.

\bibitem[Beckwith \& Sargent(1996)]{bs96} Beckwith, S.V.W., \&
Sargent, A.I. 1996, Nature, 383, 139.

\bibitem[Bessell \& Brett(1988)]{bb88} Bessell, M.S. \& Brett, J.M. 1988
\pasp, 100, 1134.

\bibitem[Chini et al.(1997)]{chini97} Chini, R., Reipurth, B.,
Ward-Thompson, D., Bally, J., Nyman, L.A., Sievers, A., \&
Billawala, Y. 1997, \apjl, 474, 135.

\bibitem[Churchwell et al.(1987)]{church87} Churchwell, E.B.,
Felli, M., Wood, D.O.S. \& Massi, M.  1987, \apj, 321, 516.

\bibitem[Cohen et al.(1981)]{cfpe81} Cohen, J.G., Frogel, J.A., Persson,
S.E. \& Elias, J.H. 1981 \apj, 249, 500.

\bibitem[Dolan \& Mathieu(1999)]{dm99} Dolan, C.J. \& Mathieu, R.D.  1999,
\aj, 118, 2409. 

\bibitem[Dougados et al.(1993)]{dlrcp93} Dougados, C., L\'{e}na, P., 
Ridgeway, S.T., Christou, J.C., \& Probst, R.G., 1993, \apj, 406, 112.

\bibitem[Duley \& Williams(1981)]{dw81} Duley, W.W., \& Williams, D.A.
1981, \mnras, 196, 269.

\bibitem[Dutrey et al.(1996)]{dutrey96} Dutrey, A., Guilloteau, S.
Duvert, G., Prato, L., Simon, M., Schuster, K. \& Menard, F.
1996, \aap, 309, 493.

\bibitem[Elias et al.(1982)]{el82} Elias, J.H., Frogel, J.A., Matthews, K.
\& Neugebauer, G.  1982, \aj, 87, 1029.

\bibitem[Felli et al.(1993)]{felli93} Felli, M. Taylor, G.B.,
Catarzi, M., Churchwell, E.B., \& Kurtz, S. 1993, \aa, 101, 127.

\bibitem[Geballe et al.(1989)]{geballe89} Geballe, T.R., Tielens, A.G.G.M.,
Allamandola, L.J., Moorhouse, A. \& Brand, P.W.J.L. 1989, \apj, 341, 278.

\bibitem[Haisch, Lada \& Lada(2000)]{hll2000} Haisch, K.E., Lada
E.A. \& Lada, C.J. 2000, \aj, (submitted March 2000).


\bibitem[Hayward \& McCaughrean(1997)]{hm97} Hayward,
T.L. \& McCaughrean, M.J. 1997 \aj,113, 346.

\bibitem[Herbig \& Terndrup(1986)]{ht86} Herbig G.H. \& Terndrup D.M.
1986, \apj, 307, 609.

\bibitem[Hillenbrand(1997)]{h97} Hillenbrand, L.A. 1997, \aj, 113, 1733.

\bibitem[Hillenbrand et al.(1998)]{hetal98} Hillenbrand, L.A., Strom,
S.E., Calvet, N., Merrill, K.M., Gatley, I., Makidon, R.B., Meyer, M.R.,
\& Skrutskie, M.F., 1998 \aj, 116, 118.

\bibitem[Johnstone, Hollenbach \& Bally(1998)]{jhb98} Johnstone, D., 
Hollenbach, D. \& Bally J. 1998 \apj, 449, 758.

\bibitem[Johnstone \& Bally(1999)]{jb99} Johnstone, D. \&
Bally J. 1999 \apjl, 510, 49.

\bibitem[Kenyon \& Hartmann(1995)]{kh95} Kenyon, S.J. \&
Hartmann, L. 1995, \apjs, 101, 117.

\bibitem[Koornneef(1983)]{koor83} Koornneef, J. 1983 \aap, 128, 84.

\bibitem[Lada \& Adams(1992)]{la92} Lada, C.J. \& Adams, F. C. 
1992, \apj, 393, 278.

\bibitem[Lada(1999a)]{clada99} Lada, C.J. 1999a, in The Origin of 
Stars and Planetary Systems, eds. C.J. Lada \& N.D. Kylafis, 
Kluwer: Dordrecht, p. 143.


\bibitem[Lada, Alves \& Lada(1996)]{lal96} Lada, C.J., Alves, J.A.,
\& Lada, E.A. 1996, \aj, 111, 1964.

\bibitem[Lada(1999b)]{elada99} Lada, E.A. 1999b, in The Origin of Stars
and Planetary Systems, eds. C.J. Lada \& N.D. Kylafis, 
Kluwer: Dordrecht, p. 441.

\bibitem[Lada \& Lada(1995)]{ll95} Lada, E.A. \& Lada, C.J. 1995,
\aj, 109, 1682.


\bibitem[Laques \& Vidal(1979)]{lv79} Laques, P. \& Vidal, J.-L. 1979,
\aa, 73, 97.

\bibitem[Marcy \& Butler(1999)]{mb99} Marcy, G.W., \& Butler,
R.P. 1999, in The Origins of Stars
and Planetary Systems, eds. C.J. Lada \& N.D. Kylafis, Kluwer:
Dordrecht, p. 681. 

\bibitem[McCullough et al.(1995)]{mcc95} McCullough, P.R. et al. 1995,
\apj, 438, 394.

\bibitem[Meyer, Calvet \& Hillenbrand(1997)]{mch97} Meyer, M.R., Calvet,
N., \& Hillenbrand, L.A. 1997, \aj, 114, 233.

\bibitem[McCaughrean \& O`Dell(1996)]{mo96} McCaughrean, M.J. \& O'Dell, 
C.R. 1996, \aj, 111, 1977.

\bibitem[McCaughrean et al.(1996)]{mrzs96} McCaughrean, M.J., Rayner, J.,
Zinnecker, H. \& Stauffer, J. 1996, in Disks and Outflows Around Young
Stars, eds. Steven Beckwith, Jakob Staude, Axel Quetz \& Antonella Natta,
Springer-Verlag: Berlin, p. 33.

\bibitem[Menard \& Bertout(1999)]{mab99} Menard, F.
\& Bertout, C. 1999, in The Origin of 
Stars and Planetary Systems, eds. C.J. Lada \& N.D. Kylafis,
Kluwer: Dodrecht, p. 341.

\bibitem[Muench et al.(2000)]{gus2000} Muench, A.A., Alves, J., Haisch,
K., Lada, C.J. \& Lada, E.A. 2000, in preparation.

\bibitem[Mundy et al.(1986)]{msbmw86} Mundy, L.G., Scoville, N., B\r{a}\r{a}th,
L.B., Masson, C.R., \& Woody, D.P. 1986, \apj, 304, L51.

\bibitem[O'Dell, Wen \& Hu(1993)]{owh93} O'Dell, C.R., Wen, Z. \& Hu, X.
1993, \apj, 410, 696.

\bibitem[O'Dell \& Wong(1996)]{ow96} O'Dell, C.R. \& Wong, S.K. 1996, \aj,
111, 846. 

\bibitem[Palla \& Stahler(1999)]{ps99} Palla, F. \& Stahler, S.W. 1999,
\apj, 525, 772.

\bibitem[Parker(1991)]{parker91} Parker, J.W.M. 1991, \pasp, 103, 243.

\bibitem[Rieke \& Lebofsky(1985)]{rl85} Rieke, G. \& Lebofsky, M. 1985 
\apj, 288, 618.

\bibitem[Roche, Aitken \& Smith(1989)]{roche89} Roche, P.F., Aitken, D.K. \& Smith,
C.H. 1989, \mnras, 236, 485.


\bibitem[Sellgren(1981)]{sell81} Sellgren, K. 1981, \apj, 245, 138.

\bibitem[Shu, Adams \& Lizano(1987)]{sal87} Shu, F.H., Adams,
F.C., \& Lizano, S. 1987, \araa 25, 23.


\bibitem[Stauffer et al.(1994)]{setal94} Stauffer, J.R., Prosser, C.F.,
Hartmann, L. \& McCaugrean M.J. 1994 \aj, 108, 1375.


\bibitem[Stetson(1987)]{Stetson87} Stetson, P.B. 1987. PASP, 99, 1915.

\bibitem[Tollestrup \& Willner(1998)]{tw98} Tollestrup, E.V. \& Willner, S.P.
1998, Proceedings of SPIE, 3354, 502.

\bibitem[Wilner et al.(2000)]{whkr00} Wilner, D.J., Ho, P.T.P.,
Kastner, J.H. \& Rodriguez. L.F. 2000, \apjl, in press.

\bibitem[Wood, Myers \& Daugherty(1994)]{wmd94} Wood, D.O.S., Myers, P.C.
\& Daugherty, D.A.  1994, \apjs, 95, 457.

\bibitem[Yu, Bally \& Devine(1997)]{ybd97} Yu, K.C., Bally, J. 
\& Devine, D. 1997, \apjl, 485, 45.



\end{thebibliography}
\end{document}